\documentclass[aps,prd,onecolumn,showpacs,groupedaddress,nofootinbib]{revtex4-2}
\usepackage{graphicx}% Include figure files
\usepackage{dcolumn}% Align table columns on decimal point
\usepackage{bm}% bold math
\usepackage{color}
\usepackage{longtable}
\usepackage{supertabular}
\usepackage[T1]{fontenc}
\usepackage{epstopdf}
\usepackage{amsmath}
\usepackage{amssymb}
\usepackage{setspace}
\usepackage{multirow}
\usepackage{booktabs}
\usepackage[section]{placeins}
\usepackage{mathrsfs}
\usepackage{hyperref}
\usepackage{xspace}

\makeatletter

\newcommand{\Rmnum}[1]{\expandafter\@slowromancap\romannumeral #1@} 
\newcommand{\bq}{\begin{equation}}
\newcommand{\eq}{\end{equation}}
\newcommand{\bqn}{\begin{eqnarray}}
\newcommand{\eqn}{\end{eqnarray}}
\newcommand{\nb}{\nonumber}

\makeatother

\def\newacronym#1#2#3{\gdef#1{\gdef#1{#2\xspace}#3 (#2)\xspace}}
\newacronym{\bh}{BH}{{\it black hole}}
\newacronym{\dm}{DM}{{\it dark matter}}
\newacronym{\isco}{ISCO}{{\it innermost stable circular orbit}}

\newcommand{\sapienza}{Dipartimento di Fisica, Sapienza Università 
	di Roma, Piazzale Aldo Moro 5, 00185, Roma, Italy}
\newcommand{\infn}{INFN, Sezione di Roma, Piazzale Aldo Moro 2, 00185, Roma, Italy}

\begin{document}
\title{A metric solution for rotating black holes embedded in dark matter halos with central spikes}

\author{Rui-Hong Yue\textsuperscript{1}}
\author{Yu-Qian Zhao\textsuperscript{2,3}}
\author{Wei-Liang Qian\textsuperscript{4,1,5,6}}\email[E-mail: ]{wlqian@usp.br (corresponding author)}

\affiliation{$^{1}$ Center for Gravitation and Cosmology, College of Physical Science and Technology, Yangzhou University, Yangzhou 225009, China}
\affiliation{$^{2}$ \sapienza}
\affiliation{$^{3}$ \infn}
\affiliation{$^{4}$ Escola de Engenharia de Lorena, Universidade de S\~ao Paulo, 12602-810, Lorena, SP, Brazil}
\affiliation{$^{5}$ School of Physical Science and Technology, Nantong University, Nantong 226019, China}
\affiliation{$^{6}$ Faculdade de Engenharia de Guaratinguet\'a, Universidade Estadual Paulista, 12516-410, Guaratinguet\'a, SP, Brazil}

\begin{abstract}
We propose an analytic metric describing rotating black holes surrounded by generic dark matter halos.
This metric is an exact solution of the field equations that incorporates a dark matter halo with a central density spike in the vicinity of the black hole.
Following the construction of the corresponding spherically symmetric solution proposed by Cardoso {\it et al.}, the rotating geometry is seeded by a mass function characterizing the dark matter distribution surrounding the black hole.
We place the inner edge of the halo at the prograde innermost stable circular orbit, a choice motivated by stable, nearly circular orbits.
For the profile studied here, the anisotropic halo joins a vacuum Kerr interior with vanishing surface stress-energy.
With time normalized at spatial infinity, the metric approaches the Minkowski metric in asymptotically Cartesian coordinates and reduces to Kerr in the absence of the halo.
In particular, it generalizes the corresponding spherically symmetric solution to the case of rotating black holes.
We discuss the physical interpretation of the model parameters and illustrate the metric by applying it to several specific gravitational systems.
\end{abstract}

%\pacs{ 03.75.Dg, 06.30.Gv, 37.25. + k, 91.10.Pp}

%\date{\today}
\date{September 29th, 2024}

\maketitle

\newpage
\section{Introduction}\label{sec1}

The nature of \dm and its distribution within galactic cores has emerged as a focal point in cosmology and astrophysics, with far-reaching implications for our understanding of the Universe~\cite{agr-dark-matter-review-02, agr-dark-matter-review-03, agr-dark-matter-review-04}. 
On the theoretical side, apart from purely analytical approaches, continuous advances into the primarily nonlinear regime of the \dm power spectrum have been fueled by $N$-body numerical simulations~\cite{agr-dark-matter-review-04}, which have indicated the presence of a central density {\it cusp}~\cite{agr-dark-matter-027,agr-dark-matter-028,agr-dark-matter-029}.
Such a hypothesized concentration of cold and collisionless \dm leads to a few immediate observational implications through astrophysical probes, as well as possible {\it direct} and {\it indirect} detections. 
Firstly, it has ignited considerable interest due to its potential to enhance high-energy radiation resulting from pronounced particle decay or annihilation processes~\cite{agr-dark-matter-034}.
As a matter of fact, our own Milky Way's center serves as a prime target for such indirect signatures of \dm~\cite{agr-dark-matter-025, agr-dark-matter-review-02}. 
The advent of cutting-edge observational tools, such as the Large Area Telescope~\cite{agr-dark-matter-031}, has ushered in an era of unprecedented precision in high-energy astronomy. 
Secondly, the distribution of \dm in galactic cores carries significant astrophysical consequences. 
The enhanced mass distribution could exert a measurable influence on the orbital dynamics of stars and other celestial bodies in the central regions of galaxies~\cite{agr-dark-matter-099, agr-dark-matter-050}. 
Meanwhile, the interplay between \dm and visible matter underscores the role of \dm in shaping galactic structure and evolution~\cite{agr-dark-matter-051, agr-dark-matter-052, agr-dark-matter-053}.
Also, the resulting weak~\cite{agr-dark-matter-035, agr-dark-matter-037, agr-dark-matter-038, agr-dark-matter-039} and strong~\cite{agr-dark-matter-030, agr-dark-matter-040} gravitational lensing might provide crucial information on \dm distribution on cosmic, cluster, and galactic scales that can be compared against theoretical calculations~\cite{agr-dark-matter-046, agr-dark-matter-047, agr-dark-matter-048}.

In particular, if a massive \bh resides at the center of the Galaxy, the distribution of the \dm is expected to feature a significant increase, giving rise to a {\it spike}\footnote{It was also known as an adiabatic cusp in the earlier literature~\cite{agr-dark-matter-023}, it is referred to as a spike~\cite{agr-dark-matter-024} to put an emphasize on the modification of the \dm halo profile owing to the central \bh.} in its density~\cite{agr-dark-matter-023, agr-dark-matter-024, agr-dark-matter-059}.
Gondolo and Silk~\cite{agr-dark-matter-024} evaluated the density profile of the \dm while a massive \bh is gradually formed by absorbing infalling matter.
The calculations are carried out by assuming that the timescale of the \dm's dynamics is shorter than that of the \bh's accretion process and, in this context, the latter is dubbed {\it adiabatic}.
In canonical perturbation theory, the orbital actions of surviving particles are conserved to adiabatic order.
As the central mass grows, particles of given actions occupy more tightly bound orbits, steepening an initially cusped halo into a spike~\cite{agr-dark-matter-024}.
The enhancement is strongest within the black hole's radius of influence, while the loss of orbits to capture depletes the innermost region.
Sadeghian {\it et al.}~\cite{agr-dark-matter-059} treated this process relativistically for a Schwarzschild black hole.
Ferrer {\it et al.}~\cite{agr-dark-matter-087} found that the spike persists for a Kerr black hole, with spin-dependent changes in density closest to the hole.
Other processes can soften the spike~\cite{agr-dark-matter-043, agr-dark-matter-044, agr-dark-matter-045}.
In the context of extreme mass-ratio inspirals, such features are expected to leave detectable imprints on both the orbital motion of the secondary object and the resulting gravitational waveform~\cite{agr-EMRI-017, agr-EMRI-018, agr-EMRI-019, agr-EMRI-040, agr-EMRI-042, agr-EMRI-043, agr-EMRI-044, agr-dark-matter-070}.
A sharp feature in the halo profile could also affect the spectral stability of the associated quasinormal modes~\cite{agr-qnm-Poschl-Teller-16, agr-qnm-Poschl-Teller-17} and give rise to echoes~\cite{agr-qnm-echoes-01, agr-qnm-echoes-review-01} in the time-domain gravitational waveforms~\cite{agr-qnm-echoes-20, agr-qnm-echoes-45, agr-qnm-instability-065}.
Moreover, the resulting image of the \bh spacetime might also suffer significant deformation and provide crucial information on \dm distribution~\cite{agr-dark-matter-056, agr-dark-matter-070, agr-strong-lensing-shadow-70}.

In realistic scenarios, the total mass of the \dm can be more significant than the central \bh.
As a result, the \bh metrics might deviate significantly from their vacuum solution counterparts.
This topic was investigated by Cardoso {\it et al.}~\cite{agr-BH-spectroscopy-024,agr-dark-matter-070} for spherically symmetric \bh{}s, where a class of solutions was developed.
The authors elaborate on a \bh metric residing in asymptotically flat spacetime, which furnishes a Hernquist-type~\cite{agr-dark-matter-021} density distribution far away from the horizon.
In the Einstein-cluster description used there, the particles move on circular orbits and the radial pressure vanishes throughout the halo.
The resulting \dm distribution is, in turn, anisotropic, and the tangential pressure is governed by the Bianchi identities.
For such an approach, it was assumed that the \dm density vanished identically at the \bh horizon.
As discussed above, analyses based on adiabatic invariants within canonical perturbation theory~\cite{agr-dark-matter-024,agr-dark-matter-059} imply an inner truncation of the available \dm phase space outside the event horizon.
Therefore, this effect should be taken into account in further studies.

Intuitively, it is physically relevant to further generalize such a metric solution to the case of spinning \bh{}s.
Nonetheless, it is a well-known challenge to derive an analytic rotating \bh{} solution for a given source~\cite{agr-bh-Kerr-05, agr-bh-Kerr-06, agr-bh-Kerr-09, agr-bh-Kerr-07}.
One widely adopted approach is the Newman-Janis algorithm~\cite{agr-Newman-Janis-01,agr-Newman-Janis-02} and its variants~\cite{agr-bh-Kerr-55}, which generate stationary and axisymmetric geometries from known static solutions through a complex coordinate transformation.
However, except for a limited class of spacetimes, the resulting metric is not guaranteed to satisfy Einstein's field equations for the corresponding matter source, and the validity and uniqueness of the complexification procedure remain subtle issues~\cite{agr-Newman-Janis-06}.
Another alternative is to construct parametrized \bh{} metrics by generalizing the Kerr solution while respecting symmetry, constants of motion, and regularity~\cite{agr-bh-Kerr-11, agr-bh-Kerr-15, agr-bh-Kerr-16, agr-bh-Kerr-30, agr-bh-Kerr-33, agr-bh-Kerr-40, agr-bh-Kerr-50}.
For instance, based on the multipole expansion, bumpy \bh{} metrics~\cite{agr-bh-Kerr-11, agr-bh-Kerr-15, agr-bh-Kerr-16} admit constants of motion, such as energy and angular momentum, and satisfy Einstein's field equations up to a few multipoles with incorrect values.
Although such constructions are desirable as model-independent parameterizations for testing alternative theories of gravity, they generally do not furnish exact solutions.
More recently, Fernandes and Cardoso generalized the model~\cite{agr-BH-spectroscopy-024} to spinning \bh{}s~\cite{agr-BH-spectroscopy-062}.
By considering a Hernquist-type profile, a Weyl-Lewis-Papapetrou class metric ansatz with axial symmetry was implemented via numerical integration, where the radial pressure was assumed to vanish identically.
In this work, following the construction of Ref.~\cite{agr-BH-spectroscopy-024}, we employ a mass function that characterizes the \dm{} distribution surrounding the \bh{} as the seed of the metric construction and derive an analytic rotating solution by solving Einstein's field equations.
The resulting metric generalizes the corresponding spherically symmetric solution to the case of spinning \bh{}s.
By imposing appropriate boundary conditions, the solution naturally accommodates a broad class of \dm{} profiles, for which the radial pressure is not required to vanish identically.

The remainder of this paper is organized as follows.
In the following section, we derive the spinning \bh metric and discuss how it falls back to various well-known cases.
In Sec.~\ref{sec3}, we explore plausible scenarios of \dm profiles featuring a spike that are consistent with existing studies.
The properties and implications of the obtained metrics are analyzed.
Numerical examples for a few well-known gravitational systems are presented in Sec.~\ref{sec5}.
Sec.~\ref{sec6} discusses how the construction relates to other halo models and summarizes our results.

\section{Spinning \bh{}s with a \dm halo}\label{sec2}

We first write down the static, spherically symmetric line element in a form equivalent to that of Cardoso {\it et al.}~\cite{agr-BH-spectroscopy-024},
\bqn
ds_0^2 &=& \left(-1+\frac{f(r)}{r^2}\right)dt^2+\frac{dr^2}{1-2m(r)/r}+r^2 d\Omega^2 , \label{MSansatz}
\eqn
where $r$ is the areal radius, $m(r)$ is the enclosed mass including the central black hole, and $d\Omega^2=d\theta^2+\sin^2\theta\,d\phi^2$.
The static stress tensor has the form $T^\mu{}_\nu=\mathrm{diag}(-\rho_s,0,p_t,p_t)$.
For a prescribed $m(r)$, Einstein's equations and stress-energy conservation give
\bqn
r(r-2m)f' &=& 2(r-m)f-2mr^2, \label{staticRadialEquation} \\
\rho_s &=& \frac{m'}{4\pi r^2},\label{asymRho} \\
p_t &=& \frac{m\rho_s}{2(r-2m)}. \label{asymPt}
\eqn
For a finite halo, normalizing the static time coordinate by $g_{tt}\to-1$ at infinity fixes the integration constant and gives
\bqn
f(r) &=& r^2\left[1-\exp\!\left(-\int_r^\infty \frac{2m(u)}{u[u-2m(u)]}\,du\right)\right]. \label{staticLapseSol}
\eqn
For the truncated profiles considered here, the enclosed mass between the horizon and the inner edge of the halo equals the black-hole mass, $m(r)=M_\mathrm{B}$.
At spatial infinity, $m(r)\to M_\mathrm{B}+M_\mathrm{H}$, where $M_\mathrm{H}$ is the total halo mass in the static solution.
In the inner vacuum region the radial equation gives $f(r)=r^2-A\,r(r-2M_\mathrm{B})$, where the constant of integration $A$ is fixed by Eq.~\eqref{staticLapseSol} when time is normalized at infinity.
With the inner time normalization used for matching below, $A=1$ and $f(r)=2M_\mathrm{B}r$ in that region.

Now, we extend this seed to finite spin by adopting the following Kerr-like {\it ansatz} in Boyer-Lindquist coordinates~\cite{book-blackhole-Chandrasekhar},
\begin{equation}
ds^2=\left(-1+\frac{f(r)}{\Sigma(r,\theta)} \right)dt^2+\frac{\Sigma(r,\theta)}{r^2+a^2-B(r)}dr^2 +\Sigma(r,\theta)d\theta^2+\sin^{2}\theta\left(r^{2}+a^{2}+\frac{f(r)a^{2}\sin^{2}\theta}{\Sigma(r,\theta)}\right)d\phi^2 -\frac{2 a\sin^2\theta f(r)}{\Sigma(r,\theta)}dt d\phi, \label{Mansatz}
\end{equation}
where $a$ is the spin of the \bh,
\begin{equation}
    \Sigma(r,\theta)\equiv\Sigma =r^2+a^2 \cos^2\theta
\end{equation}
and $f(r)$ and $B(r)$ are two radial functions.
Setting $a=0$ gives $\Sigma=r^2$, and Eq.~\eqref{Mansatz} reduces directly to Eq.~\eqref{MSansatz} with the same $f(r)$ and $B(r)=2rm(r)$.
At finite spin we keep the radial relation between $f$ and $B$ inherited from the seed, as made explicit in Eq.~\eqref{relfB}, and prescribe $m(r)$ separately.
We match the halo to a vacuum Kerr interior at the radius of the prograde innermost stable circular orbit, $r_{\rm ISCO}(a)$.
At spatial infinity we require the halo density and stresses to decay.
In particular, $p_r$ vanishes for $r\leq r_{\rm ISCO}$ and tends to zero as $r\to\infty$, as stated in Eqs.~\eqref{prCond1} and~\eqref{prCond2}.
If the mass profile also satisfies $m'(r_{\rm ISCO}^{+})=0$, the bulk density and tangential stresses vanish on approaching the inner edge from the halo side.
The matching data are derived in Sec.~\ref{sec3}, and Fig.~\ref{illusDKModel01} illustrates the resulting density and radial pressure.

By substituting the above metric ansatz into the Einstein field equation 
\bqn
G_{\mu\nu} = 8\pi T_{\mu\nu} ,
\eqn
where one assumes $c=G=1$. 

To assess the \dm profile, we project the stress-energy tensor onto the following orthonormal tetrad:
\bqn
e^\mu_t&=&\left(
\frac{a^2+r^2}{\Sigma^{1/2}(r^2+a^2-f)^{1/2}}, 0, 0, \frac{a}{\Sigma^{1/2}(r^2+a^2-f)^{1/2}}\right), \nonumber \\
e^\mu_r&=&\left(0, \frac{(r^2+a^2-B)^{1/2}}{\Sigma^{1/2}}, 0, 0\right),  \nonumber \\
e^\mu_\theta&=&\left(0, 0, \frac{1}{\Sigma^{1/2}}, 0\right),  \nonumber \\
e^\mu_\phi&=&\left(-\frac{a\sin\theta}{\Sigma^{1/2}}, 0, 0, -\frac1{\Sigma^{1/2}\sin\theta}\right),
\eqn
which satisfies
\begin{equation}
e^\mu_a\eta^{ab}e^\nu_b=g^{\mu\nu}.
\end{equation}
The above tetrad corresponds to a stationary orthonormal observer with respect to the background black-hole spacetime.
Consequently, the quantities obtained below are the local projections of the stress-energy tensor measured by this observer, and need not coincide with the principal energy density and principal pressures of the matter source.

The corresponding tetrad projections are
\bqn
\rho &=& e^\mu_t T_{\mu\nu}e^\nu_t = \frac{1}{8\pi}\frac{a^4\cos^2\theta\left( B' r -3f+2B \right)+a^2\cos^2\theta\left( B'r^3-f \cdot B' r-2f r^2+Br^2\right)}{\left( r^2+a^2\cos^2\theta \right)^3\left( r^2+a^2-f \right)}\nonumber\\
&&+\frac{1}{8\pi}\frac{a^2\cos^2\theta\left( 3f^2 -2fB\right)+\left(B'r^3-fr^2 \right)a^2+B'r^5-fB'r^3-Br^4+fBr^2}{\left( r^2+a^2\cos^2\theta \right)^3\left( r^2+a^2-f \right)},\label{rhoDef}\\
p_r &=&e^\mu_r T_{\mu\nu}e^\nu_r =\frac{1}{8\pi}\frac{a^4\cos^2\theta \left(  f'r-f \right)+a^2 \cos^2\theta \left(f'r^3 -2f r^2-B f'r+B r^2+f^2\right)}{\left( r^2 +a^2\cos^2\theta\right)^3\left( f-r^2-a^2 \right)}\nonumber \\
& &+\frac{1}{8\pi}\frac{a^2\left(  f'r^3-fr^2 \right)+r^2\left(  f'r^3-2fr^2-B f'r+Br^2+Bf \right)}{\left( r^2 +a^2\cos^2\theta\right)^3\left( f-r^2-a^2 \right)} ,\label{prDef} \\
p_\theta&=& e^\mu_\theta T_{\mu\nu}e^\nu_\theta =\frac{1}{32\pi}\frac{a^4\cos^4{\theta}\ F+a^2\cos^2{\theta}\ G+H}{\Sigma^3\left(r^2+a^2-f\right)^2} ,\label{pthetaDef}\\
p_\phi &=& e^\mu_\phi T_{\mu\nu}e^\nu_\phi =p_\theta -\frac{1}{8\pi}\frac{2 \sin^2{\theta}\left(f(r)-B(r)\right) a^2r^2}{\Sigma^3\left(r^2+a^2-f\right)}, \label{pphiDef}
\eqn
where
\bqn
F &=&-2\left(r^2+a^2-f\right)\left(r^2+a^2-B\right) f"+\left(B-r^2-a^2\right){f'}^2 \nonumber\\
& &+(r^2+a^2-f)B'f'+2r(r^2+a^2+f-2B)f' +2r(f-r^2-a^2)B'+4(B-f)(f-a^2) ,\nonumber\\
G &=& -4r^2(r^2+a^2-B)(r^2+a^2-f)f"+2r^2(B-r^2-a^2)(f')^2+2r^2(r^2+a^2-f)B'f'+8r^3(r^2+a^2-B)f'\nonumber\\
& &+4r(r^2+a^2-f)(a^2-B)f'+4r^2(f-r^2-a^2)(B+rB')+8r^4(B-f)-4(r^2+a^2-f)(a^2-f)f ,\nonumber
\\
H &=&-2r^4(f-r^2-a^2)(B-r^2-a^2)f"+r^4(B-r^2-a^2)(f')^2-r^4(f-r^2-a^2)B'f' \nonumber\\
& &+\left(4r^3(r^2+a^2-f)(r^2+a^2-B)+2r^5(r^2+a^2-B)+2r^5(f-B)\right)f' \nonumber\\
& &+2r^5(f-r^2-a^2)B'+4r^6(B-f)-4r^2(r^2+a^2-f)f(r^2+a^2-B),\nonumber
\eqn
where $f\equiv f(r)$, $B\equiv B(r)$, and the prime represents differentiation with respect to the radial coordinate.

Specifically, one finds that the stress-energy tensor takes the following anisotropic form in the orthonormal frame:
\bqn\label{Tab}
T_{\hat{a}\hat{b}}=
\begin{pmatrix}
\rho & 0 & 0 & T_{\hat t\hat\phi}\\
0 & p_r & 0 & 0\\
0 & 0 & p_\theta & 0\\
* & 0 & 0 & p_\phi
\end{pmatrix},
\eqn
where
\bqn
T_{\hat t\hat\phi}
=e^\mu_t T_{\mu\nu}e^\nu_\phi=
-\frac{1}{8\pi}\frac{a \sin\theta[rB'(r^2+a^2-f)-rf'(r^2+a^2-B)+2(a^2-f)(B-f)] }{2 (r^2+a^2 \cos^2\theta)^2(r^2+a^2-f)^{3/2}} ,\label{tPhiProjection}
\eqn
while the remaining off-diagonal components vanish identically as a consequence of the circularity condition satisfied in the Boyer-Lindquist (Papapetrou) coordinates.

The nonvanishing off-diagonal component $T_{\hat t\hat\phi}$ represents the azimuthal momentum density of the dark matter measured by the stationary orthonormal observer associated with the above tetrad.
Its presence indicates that this observer is not comoving with the rotating matter distribution.
Consequently, the quantities $\rho$, $p_r$, $p_\theta$, and $p_\phi$ obtained above should be interpreted as the energy density and normal stresses measured by the stationary observer, rather than the principal energy density and principal pressures of the matter source.

To identify the latter, one introduces a local Lorentz boost in the $(\hat t,\hat\phi)$ plane.
As shown in Appx.~\ref{appB}, provided that
\begin{equation}\label{condTtphi}
(\rho+p_\phi)^2>4T_{\hat t\hat\phi}^{\,2},
\qquad
\rho+p_\phi>0,
\end{equation}
the boost velocity is real and subluminal, and a local rest frame exists in which $T_{\hat t'\hat\phi'}=0$.
In that frame the diagonal entries of $T_{\hat a\hat b}$ give the principal energy density and azimuthal pressure.
The radial and polar pressures are unchanged by the boost.

For the metric in Eq.~\eqref{Mansatz} to recover the static seed, its radial pressure must obey
\bqn
\left.p_r\right|_{a=0}=0 .
\label{pr0Cond}
\eqn
Evaluating Eq.~\eqref{prDef} at $a=0$ gives
\bqn
f'(r)r^3-2f(r)r^2-B(r){f'(r)}r+B(r)r^2+B(r)f(r) = 0, \label{fcond}
\eqn
which implies 
\bqn
f'(r)=\frac{B(r)r^2+B(r)f(r)-2f(r)r^2}{r\left( B(r)-r^2 \right)} .\label{relfB}
\eqn
Substituting $B=2rm$ in Eq.~\eqref{relfB} gives precisely the static equation for $f$ in Eq.~\eqref{staticRadialEquation}.
Using this same relation for each $a$ is our prescription for the rotating extension.
The zero-spin condition alone leaves this continuation undetermined.
With this choice the density and radial pressure take the form
\bqn
\rho &=& \frac{rB'-B}{8\pi\Sigma^2}
+\frac{a^2(B-f)\left[r^2+(2r^2+3a^2-3f)\cos^2\theta\right]}
{8\pi\Sigma^3(r^2+a^2-f)}, \label{rhoDef2}\\
p_r &=& \frac{a^2(B-f)\left[r^4+\bigl(a^2r^2+f(r^2-B)\bigr)\cos^2\theta\right]}
{8\pi(r^2-B)\Sigma^3(r^2+a^2-f)}. \label{prDef2}
\eqn
With $B=2rm$, these expressions give $\rho=m'/(4\pi r^2)$ and $p_r=0$ at $a=0$.
Eqs.~\eqref{pthetaDef} and~\eqref{pphiDef} then give $p_\theta=p_\phi=p_t$ as in Eq.~\eqref{asymPt}.

To proceed, we introduce an auxiliary metric function $q\equiv q(r)$
\bqn
f(r) = 2 q(r) r , \label{qmassFunc}
\eqn
based on the metric function $f(r)$.
%and denote the total mass of the \dm halo by
%\bqn
%M_\mathrm{Halo} = \int \rho r^2drd\Omega .\label{haloMass}
%\eqn
With Eq.~\eqref{qmassFunc}, Eq.~\eqref{relfB} becomes
\begin{equation}
B(r)=\frac{2r\left( q'(r)r-q(r) \right)}{2 q'(r)-1} .\label{cond1}
\end{equation}
Therefore, the metric is entirely determined by Eqs.~\eqref{qmassFunc} and~\eqref{cond1} for a given $q(r)$.

We write $B(r)$ in terms of the prescribed mass function introduced in Eq.~\eqref{MSansatz},
\bqn
B(r) = 2 m(r) r , \label{massFunc}
\eqn
which implies
\bqn
m(r)=\frac{q'(r)r-q(r)}{2 q'(r)-1}. \label{BmassFunc}
\eqn
The above relation can be integrated analytically to give
\bqn
q(r)=\frac12\left[r-\sigma \exp\left(-\int\frac{dr}{2m(r)-r}\right)\right] ,\label{BmassFuncInv}
\eqn
where $\sigma=\pm 1$ governed by the size of $q$.
It is apparent that, in the vacuum Schwarzschild limit, one takes $\sigma=+1$, and the functions $q(r)$ and $m(r)$ are both identical to the \bh mass, $q(r)=m(r)=M_\mathrm{B}$.
As elaborated below, Eq.~\eqref{massF_cond_Qf} indicates that $\sigma=+1$ continues to be valid in the presence of \dm.

For a prescribed $m(r)$, Eq.~\eqref{BmassFuncInv} determines $q(r)$ up to one integration constant, which is fixed by matching at the inner edge in Sec.~\ref{sec3}.

At this stage, the matching condition in Eq.~\eqref{pr0Cond} alone does not guarantee a physically plausible scenario.
An additional requirement is that the density, and consequently the pressure, vanish below a certain radius within the spike.
As discussed in the Introduction, the restriction of the available phase space near the \bh naturally leads to an inner truncation of the matter distribution.
Identifying the inner edge with the prograde \isco is a modeling assumption for a population on stable, nearly circular orbits.
Ref.~\cite{agr-EMRI-056} uses the ISCO as the endpoint of a quasi-circular inspiral.
In Schwarzschild spacetime, bound eccentric orbits can have pericenters below the ISCO at $6M_\mathrm{B}$ and arbitrarily close to the marginally bound radius $4M_\mathrm{B}$ from above~\cite{agr-dark-matter-059}.
In Kerr spacetime, capture depends on the orbital energy, axial angular momentum, and Carter constant~\cite{agr-dark-matter-087}.
For a halo populated by such orbits, one may prescribe a smaller inner radius $r_{\rm in}$ for the mass profile and use it as the lower limit in Eq.~\eqref{BmassFuncFormal}, provided the matching surface is timelike and the radial solution is regular.
For the adopted cutoff, the inner pressure condition is
\bqn
p_r(r\leq r_{\rm ISCO})=0 .
\label{prCond1}
\eqn
Meanwhile, the radial pressure must also vanish at spatial infinity:
\bqn
\lim\limits_{r\to\infty}p_r(r)=0 .
\label{prCond2}
\eqn
These conditions will be examined in detail in the next section.

In the spherical limit, Cardoso {\it et al.}~\cite{agr-BH-spectroscopy-024} use $f$ for the function multiplying $-dt^2$.
Denoting their function by $f_\ast(r)$, we have $g_{tt}=-f_\ast(r)$, so comparison with Eq.~\eqref{MSansatz} in the same time coordinate gives
\bqn
f_\ast(r) &=& 1-\frac{f(r)}{r^2},\qquad f(r)=r^2[1-f_\ast(r)]. \label{staticLapseMap}
\eqn
With the identification in Eq.~\eqref{staticLapseMap}, the radial equation in Eq.~\eqref{staticRadialEquation} takes their form $r f_\ast'/(2f_\ast)=m/(r-2m)$.

For $a\ne0$, the special choice $f=B$ makes $p_r$ vanish identically in Eq.~\eqref{prDef2}.
Eq.~\eqref{relfB} then gives $rf'=f$.
Writing the integration constant as $2M_\mathrm{B}$ yields the vacuum Kerr solution,
\bqn
f(r) = B(r) = 2M_\mathrm{B} r .\label{ConFallBack}
\eqn
This is the interior solution used for matching in Sec.~\ref{sec3}.

Before closing this section, we remark that an explicit comparison between the present metric and those obtained using the Newman-Janis algorithm and its alternatives reveals substantial differences in their underlying constructions.
For clarity of presentation, we relegate the detailed discussion to Appx.~\ref{appA}.

\section{Plausible \dm profile with a spike}\label{sec3}

To ensure the condition Eq.~\eqref{prCond1}, a sufficient condition for a vanishing radial pressure at the \isco radius $r_\mathrm{ISCO}$ can be derived from Eq.~\eqref{prDef2}: 
\bqn
\lim\limits_{r\to r_\mathrm{ISCO}+}\left(B(r)-f(r)\right) = 0 .\label{fBcon1}
\eqn

For $r\le r_\mathrm{ISCO}$, the metric is essentially Kerr, namely,
\bqn
\left.B(r)=f(r)\right|_{r \le r_\mathrm{ISCO}} = 2M_\mathrm{B} r .\label{fBCon2}
\eqn
 
Taking into account Eq.~\eqref{cond1}, Eq.~\eqref{fBcon1} implies that
\bqn
\lim\limits_{r\to r_\mathrm{ISCO}+} q'(r)(2q(r)-r) =  0 ,
\eqn
namely, either
\bqn
\lim\limits_{r\to r_\mathrm{ISCO}+} q'(r) =  0 ,\label{proSce1}
\eqn
or
\bqn
\lim\limits_{r\to r_\mathrm{ISCO}+} 2q(r) =  {r_\mathrm{ISCO}} .\label{proSce2}
\eqn
However, Eq.~\eqref{fBCon2} implies 
\bqn
\lim\limits_{r\to r_\mathrm{ISCO}-} q(r)=M_\mathrm{B} .\label{isoMF}
\eqn
Owing to Eq.~\eqref{BmassFuncInv}, $q(r)$ is continuous at $r=r_\mathrm{ISCO}$.
For $r_\mathrm{ISCO}\ne 2M_\mathrm{B}$, Eq.~\eqref{isoMF} excludes the branch in Eq.~\eqref{proSce2}.
By substituting the remaining condition Eq.~\eqref{proSce1} into the definition Eq.~\eqref{BmassFunc}, one finds
\bqn
q(r=r_\mathrm{ISCO}) = m(r=r_\mathrm{ISCO})=M_\mathrm{B}  ,\label{proSce1ISO}
\eqn
which is rather desirable viewed together with Eq.~\eqref{isoMF}.
The radial coordinate $r=r_\mathrm{ISCO}$ can be used to fix the constant of integration when one evaluates $q(r)$ using numerical integral for a given mass function $m(r)$.

On the other hand, at spatial infinity, as the \dm dilutes, one expects both the density and pressure to vanish identically.
Therefore, using Eq.~\eqref{asymRho}, a necessary condition for an asymptotically flat spacetime is
\bqn
\lim\limits_{r\to \infty} \frac{m'(r)}{r^2} = 0 . \label{assymMF}
\eqn

Regarding the general requirement of the mass function, one can elaborate a bit further.
Firstly, since $m(r)$ increases monotonically with the radial coordinate, $m'$ is positive definite.
Using Eqs.~\eqref{cond1} and~\eqref{massFunc}, the metric ansatz indicates that the event horizons of the metric governed by the roots of the function
\bqn
r^2+a^2-B(r) = r^2+a^2-2m(r)r = r^2+a^2-\frac{2r\left( q'(r)r-q(r) \right)}{2 q'(r)-1} = 0 .\label{EqRhKerrDM}
\eqn
For $r<r_\mathrm{ISCO}$, the two roots of Eq.~\eqref{EqRhKerrDM} possess the form
\bqn
r_h^\mathrm{\pm} = M_\mathrm{B} \pm \sqrt{M^2_\mathrm{B} - a^2} ,\label{rhKerr}
\eqn
For $r \ge r_\mathrm{ISCO}$, one needs to ascertain that the potential roots are smaller than the \isco radius and thus become irrelevant.
It is straightforward to show that a sufficient condition for any root to be smaller than $r_\mathrm{ISCO}$ is%, as the latter leads to the l.h.s. of Eq.~\eqref{EqRhKerrDM} being positive definite
\bqn
r > 2m(r) . \label{massF_cond}
\eqn
The condition Eq.~\eqref{massF_cond} is rather intuitive as it can be physically interpreted as that the radial coordinate must be larger than the corresponding Schwarzschild radius, namely, $r_\mathrm{Sch}\equiv 2m(r)$.
Besides, independent of any specific form of $q(r)$, Eq.~\eqref{EqRhKerrDM} can be integrated once to give
\bqn
r-2q(r) = C e^{-r^2/2 a^2} ,\label{invCon}
\eqn
where $C$ is the constant of integration.
It is immediately recognized that this equation falls back to that which governs the Schwarzschild radius at the limit $a\to 0$.
Besides, for $a\ne 0$, the l.h.s. of Eq.~\eqref{invCon} is either positive definite or negative definite.
We assert that $C$ must be positive for a physically relevant metric solution.
This is because the assumption $C<0$ would imply that the entire spacetime beyond the root in question is acausal. 
Given that the requirement that there is no more horizon for the relevant region of radial coordinate is guaranteed by 
\bqn
r^2 + a^2 - \frac{2r\left( q'(r)r-q(r) \right)}{2 q'(r)-1} > 0 ,\label{inEquaCond}
\eqn
a necessary condition is
\bqn
r > 2q(r) . \label{massF_cond_Qf}
\eqn
%In practice, one can use either Eq.~\eqref{massF_cond} or~\eqref{massF_cond_Qf} as a sufficient constraint for the proposed model.

Now, it is a good place to pause and summarize the mathematical problem so far, ensuring a physically relevant \dm profile.
One starts with a mass function $m(r)$ that satisfies Eqs.~\eqref{proSce1ISO} and~\eqref{assymMF}.
The radial mass profile can be supplied by a halo model or constrained by observations.
Subsequently, the metric function $q(r)$ can be derived by Eq.~\eqref{BmassFuncInv}, whose constant of integration must be chosen to satisfy the boundary conditions Eqs.~\eqref{proSce1} and~\eqref{isoMF}.
Given the functions $m(r)$ and $q(r)$ and the definitions given by Eqs.~\eqref{qmassFunc} and~\eqref{massFunc}, the metric Eq.~\eqref{Mansatz} is determined.
Further, one must guarantee the resulting metric has vanishing radial pressure $p_r$ Eq.~\eqref{prCond1} at $r=r_\mathrm{ISCO}$ and Eq.~\eqref{prCond2} at spatial infinity.

Before proceeding further, we note that since $p_r\ne p_{\theta, \phi} (\ne 0)$ in general, the \dm distribution is anisotropic.
For $r\le r_\mathrm{ISCO}$, $\rho = p_{\theta, \phi} = T_{\hat{t}\hat{\phi}} = 0$ as matter and pressure vanish identically, as expected.

We apply the Israel junction conditions~\cite{agr-collapse-thin-shell-03} at the timelike inner edge $r=r_{\rm in}$, with $r_{\rm in}=r_{\rm ISCO}$ in the example considered here.
Appx.~A of our earlier study of black hole shadows~\cite{agr-strong-lensing-shadow-35} illustrates the same junction calculation for a slowly rotating massive shell with a nonzero jump in the extrinsic curvature.
If $m(r)-M_\mathrm{B}$ vanishes quadratically at the edge from the halo side, $m(r)$ is continuously differentiable there and Eq.~\eqref{asymRho} gives a static density that rises linearly from zero.
On each side we orient the unit normal $n^\mu=\sqrt{g^{rr}}\,\delta^\mu_r$ from the Kerr interior toward the halo and define $[X]=X_{\rm halo}-X_{\rm Kerr}$.
In the common tangential coordinates $y^i=(t,\theta,\phi)$, Eq.~\eqref{Mansatz} gives the induced metric $h_{ij}=g_{ij}$ and the extrinsic curvature $K_{ij}=\tfrac12\sqrt{g^{rr}}\,\partial_r g_{ij}$.
The Kerr interior in Eq.~\eqref{fBCon2} and the matching data in Eqs.~\eqref{proSce1} and~\eqref{proSce1ISO} give $m=q=M_\mathrm{B}$ and $q'=0$ on both sides at the edge.
Hence $B=f=2M_\mathrm{B}r_{\rm in}$ and $f'=2M_\mathrm{B}$ there, so $[h_{ij}]=[g^{rr}]=0$ and the remaining jump is
\bqn
[K_{ij}]\,dy^i dy^j &=& \left.\frac{\sqrt{g^{rr}}}{2\Sigma}\right|_{r_{\rm in}}[f']\bigl(dt-a\sin^2\theta\,d\phi\bigr)^2=0 . \label{junctionCurvatureJump}
\eqn
The surface stress tensor therefore vanishes by the Israel relation $8\pi S_{ij}=h_{ij}[K]-[K_{ij}]$, where $[K]=h^{ij}[K_{ij}]$.

Motivated by the results for vacuum \bh solution, it is rather attempting to assume $q'(r)\to 0$ at spatial infinity, and therefore Eq.~\eqref{BmassFunc} leads to
\bqn
\lim\limits_{r\to \infty} q(r)  \overset{?}{=} \lim\limits_{r\to \infty} m(r) .\label{proSce1Inf}
\eqn
Subsequently, this implies that at large radial coordinate, $q(r)$ converges and asymptotically approaches the mass function $m(r)$.
A separate idealization concentrates the halo mass in a $\delta$-function shell at $r_\mathrm{SP}\ge r_\mathrm{ISCO}$.
Specifically we have
\begin{equation}
m(r) = M_\mathrm{B}+M_\mathrm{H}\Theta(r-r_\mathrm{SP}) ,
\label{massF_deltaSP}
\end{equation}
where $\Theta(x)$ is the step function.
Choosing the constant Kerr branch of Eq.~\eqref{BmassFunc} in each vacuum region gives
\begin{equation}
q(r) = 
\begin{cases}
   M_\mathrm{B}, &  r < r_\mathrm{SP}, \\
   M_\mathrm{B}+M_\mathrm{H}, &  r \ge r_\mathrm{SP}, 
\end{cases}
\label{qFunc_deltaSP}
\end{equation}
where each branch is written in its own Kerr coordinates.
Joining the branches in Eq.~\eqref{qFunc_deltaSP} across $r_\mathrm{SP}$ requires a separate junction calculation.
Although Eq.~\eqref{qFunc_deltaSP} is intuitive and manifestly satisfies the relation Eq.~\eqref{proSce1Inf}, as we argue below, this is actually {\it not} the most general case in the presence of a \dm halo.

For a given monotonically increasing mass function with the boundary value Eq.~\eqref{proSce1ISO}, the integration Eq.~\eqref{BmassFuncInv} formally gives
\bqn
q(r)=\frac12\left[r-\left(r_\mathrm{ISCO}-2M_\mathrm{B}\right) \exp\left(-\int_{r_\mathrm{ISCO}}^r\frac{dr'}{2m(r')-r'}\right)\right] .\label{BmassFuncFormal}
\eqn
Eq.~\eqref{BmassFuncFormal} incorporates the inner matching conditions in Eqs.~\eqref{proSce1} and~\eqref{isoMF} and gives $q(r)$ by numerical quadrature for a prescribed $m(r)$.
Differentiating Eq.~\eqref{BmassFuncFormal} gives $\partial_r\ln[(r-2q)/r]=2m/[r(r-2m)]$.
For a regular matched solution with finite $m(\infty)$, the right-hand side is $O(r^{-2})$ and is integrable at infinity, so the inner matching fixes the limit
\bqn
\lim\limits_{r\to \infty}\frac{1}{r}\left(r_\mathrm{ISCO}-2M_\mathrm{B}\right) \exp\left(-\int_{r_\mathrm{ISCO}}^r\frac{dr'}{2m(r')-r'}\right) = \mathcal{C} ,\label{assLinearC}
\eqn
where $\mathcal{C}$ is a positive constant.
Specifically, as $m(r)$ increases from $M_\mathrm{B}$ to $M_\mathrm{B}+M_\mathrm{H}$, the constant $\mathcal{C}$ must lie in the range
\bqn
1 \le \mathcal{C} \le \frac{r_\mathrm{ISCO}-2M_\mathrm{B}}{r_\mathrm{ISCO}-2\left(M_\mathrm{B}+M_\mathrm{H}\right)}  ,\nb
\eqn
if $r_\mathrm{ISCO} > 2\left(M_\mathrm{B}+M_\mathrm{H}\right)$, otherwise
\bqn
1 \le \mathcal{C} < +\infty  .\nb
\eqn
The constant exterior Kerr branch in Eq.~\eqref{qFunc_deltaSP} has no linear term in its own Kerr coordinates.
For the matched halo solution, Eq.~\eqref{assLinearC} gives $q(r)/r\to(1-\mathcal C)/2$, and the radial pressure in Eq.~\eqref{prDef2} tends to zero.

With $f=2rq$ from Eq.~\eqref{qmassFunc} and $\Sigma/r^2\to1$, Eqs.~\eqref{Mansatz} and~\eqref{assLinearC} give $g_{tt}\to-\mathcal C$ in the time coordinate $t$ fixed by the inner Kerr matching.
With $T=\sqrt{\mathcal C}\,t$, the stationary Killing vector $\partial_T$ has squared norm $-1$ at infinity, and the time components transform as
\bqn
g_{TT} &=& \frac{g_{tt}}{\mathcal C}\longrightarrow -1, \label{asymptoticTimeNorm} \\
g_{T\phi} &=& \frac{g_{t\phi}}{\sqrt{\mathcal C}} . \label{normalizedTimeAzimuth}
\eqn
In the static limit the coefficient of the inner vacuum lapse is thus $A=1/\mathcal C$ when time is normalized as in Eq.~\eqref{staticLapseSol}.

To gain further intuition, we proceed by considering a more explicit example.
Following the strategy of Cardoso {\it et al.}~\cite{agr-BH-spectroscopy-024}, we prescribe an analytic mass function.
To represent a halo with an inner cutoff and an adjustable slope, we take
\begin{equation}
m(r) = 
\begin{cases}
%   M_\mathrm{B}, &  r \le 2M_\mathrm{B}, \\
%   M_\mathrm{B}, &  2M_\mathrm{B} < r < r_\mathrm{ISCO}, \\
   M_\mathrm{B}, &  r < r_\mathrm{ISCO}, \\
   M_\mathrm{B}+M_\mathrm{H}\left(\frac{r}{r_\mathrm{SP}+r}\right)^{\alpha}\left(1-\frac{r_\mathrm{ISCO}}{r}\right)^2, &  r \ge r_\mathrm{ISCO},
\end{cases}
\label{massF_ansatz}
\end{equation}
where $\alpha>0$ sets the intermediate density slope and $r_{\rm SP}$ sets the scale of the transition to the outer profile.
In the static limit, $r_{\rm ISCO}=6M_\mathrm{B}$.
Differentiating Eq.~\eqref{massF_ansatz} and using $4\pi r^2\rho_s=m'$ from Eq.~\eqref{asymRho} gives, for $r>r_{\rm ISCO}$,
\bqn
\rho_s(r) &=& \frac{M_\mathrm{H}}{4\pi r^3}
\left(\frac{r}{r+r_{\rm SP}}\right)^\alpha
\left(1-\frac{r_{\rm ISCO}}{r}\right)
\left[\frac{\alpha r_{\rm SP}}{r+r_{\rm SP}}\left(1-\frac{r_{\rm ISCO}}{r}\right)+\frac{2r_{\rm ISCO}}{r}\right]. \label{rhoStaticSpike}
\eqn
The density vanishes at $r_{\rm ISCO}$ and throughout the inner vacuum region.
For $r_{\rm ISCO}\ll r\ll r_{\rm SP}$, the factor $[r/(r+r_{\rm SP})]^\alpha$ approaches $(r/r_{\rm SP})^\alpha$ and the square bracket approaches $\alpha$.
Hence
\bqn
\rho_s(r) &=& \frac{\alpha M_\mathrm{H}}{4\pi r_{\rm SP}^{\alpha}}\,r^{-(3-\alpha)}
\left[1+O\!\left(\frac{r_{\rm ISCO}}{r}\right)+O\!\left(\frac{r}{r_{\rm SP}}\right)\right]. \label{rhoSpikeIntermediate}
\eqn
Thus $\alpha=2$ gives the inner $r^{-1}$ slope of the Navarro-Frenk-White (NFW) cusp~\cite{agr-dark-matter-026,agr-dark-matter-027}, and $\alpha=2/3$ gives the $r^{-7/3}$ spike predicted for adiabatic growth in an initially $r^{-1}$ cusp~\cite{agr-dark-matter-024}.
These slopes apply between the two radial scales.
Extending the intermediate power law to infinity would give an enclosed halo mass growing as $r^\alpha$.
For $r\gg\max(r_{\rm ISCO},r_{\rm SP})$, Eq.~\eqref{rhoStaticSpike} gives $\rho_s(r)=M_\mathrm{H}(\alpha r_{\rm SP}+2r_{\rm ISCO})/(4\pi r^4)+O(r^{-5})$ for every $\alpha>0$.
The $r^{-4}$ falloff gives a finite total halo mass.
The full NFW profile falls as $r^{-3}$ in its outer halo~\cite{agr-dark-matter-026,agr-dark-matter-027}.
The spherical mass parametrization of Ghosh {\it et al.}~\cite{agr-dark-matter-100} includes the halo contribution in Eq.~\eqref{massF_ansatz} for $n_1=\alpha$, $n_2=n_3=1$, a quadratic inner taper, and an outer cutoff $r_c\to\infty$.
For the spin-dependent inner edge we use the standard prograde Kerr ISCO radius~\cite{book-blackhole-Chandrasekhar},
\begin{equation}\label{defISCO}
r_{\mathrm{ISCO}} = M_\mathrm{B} \left[ 
3 + Z_2 - 
\sqrt{(3 - Z_1)(3 + Z_1 + 2 Z_2)}
\right],
\end{equation}
with
\begin{equation}
Z_1 = 1 + \left(1 - \frac{a^2}{M_\mathrm{B}^2}\right)^{1/3}
\left[
\left(1 + \frac{a}{M_\mathrm{B}}\right)^{1/3}
+
\left(1 - \frac{a}{M_\mathrm{B}}\right)^{1/3}
\right],
\end{equation}
and
\begin{equation}
Z_2 = \sqrt{
3 \frac{a^2}{M_\mathrm{B}^2} + Z_1^2
}.
\end{equation}
Eq.~\eqref{defISCO} gives the prograde equatorial ISCO and reduces to $r_{\rm ISCO}=6M_\mathrm{B}$ in the Schwarzschild limit.
It is apparent that the above proposal is slightly revised with respect to the recipe in~\cite{agr-BH-spectroscopy-024, agr-dark-matter-070}.
There, the event horizon is located at the Schwarzschild radius $r_h^0=2M_\mathrm{B}$, where the \dm density is assumed to vanish.
In our example the nearly circular population sets the inner edge at $r_\mathrm{ISCO}>r_h$, as discussed in Sec.~\ref{sec2}.
It is not difficult to see Eq.~\eqref{massF_ansatz} is a solution that meets this requirement together with the boundary conditions Eqs.~\eqref{proSce1ISO} and~\eqref{assymMF}.
Therefore, the proposed profile can be viewed as a natural generalization of the scenario proposed in~\cite{agr-BH-spectroscopy-024}.

\begin{figure}[ht]
    \centering
    \begin{minipage}{0.4\textwidth}
        \centering
        \includegraphics[width=1.0\textwidth,height=0.75\textwidth]{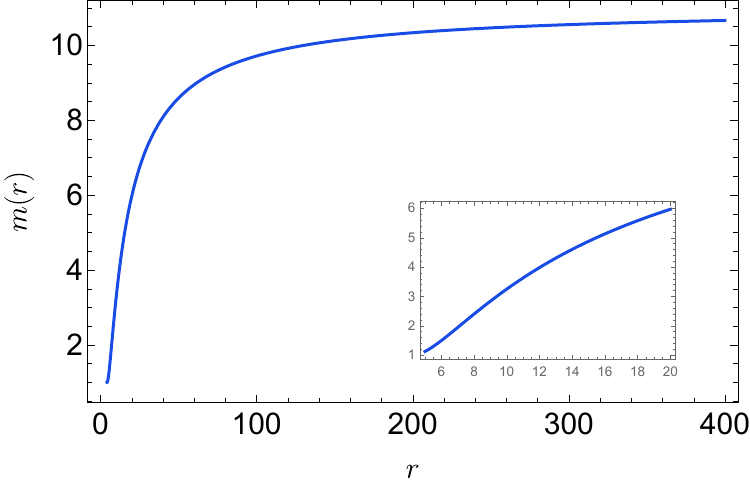}
    \end{minipage}
    \begin{minipage}{0.4\textwidth}
        \centering
        \includegraphics[width=1.0\textwidth,height=0.75\textwidth]{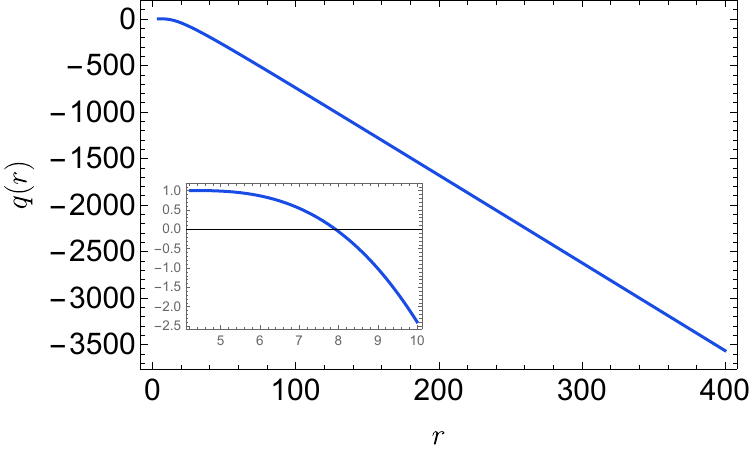}
    \end{minipage}
    \begin{minipage}{0.4\textwidth}
        \centering
        \includegraphics[width=1.0\textwidth,height=0.75\textwidth]{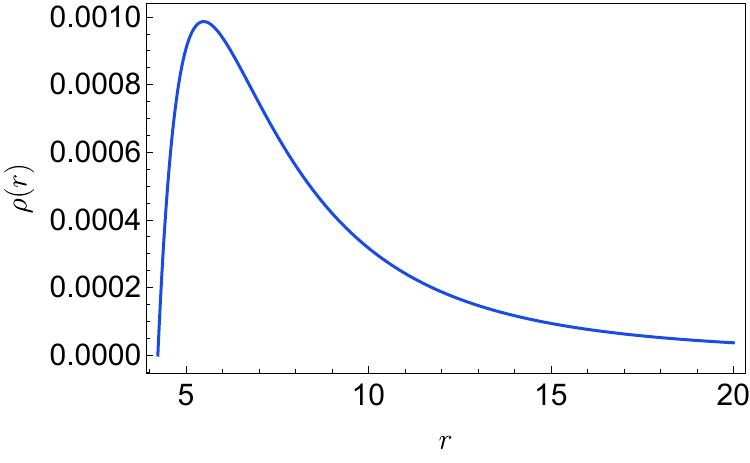}
    \end{minipage}
    \begin{minipage}{0.4\textwidth}
        \centering
        \includegraphics[width=1.0\textwidth,height=0.75\textwidth]{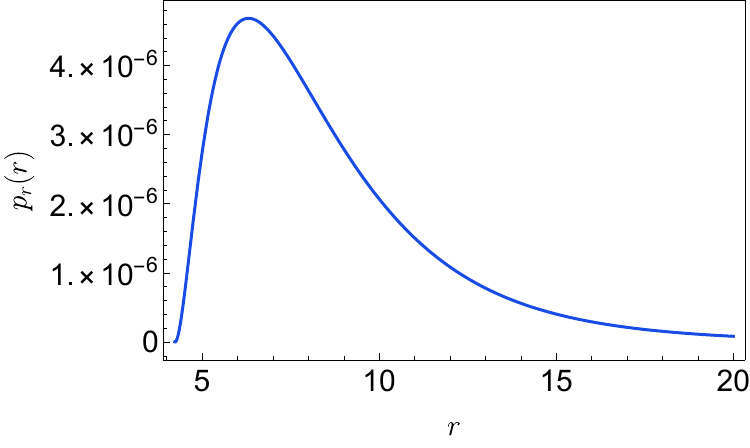}
    \end{minipage}
\renewcommand{\figurename}{Fig.}   
\caption{An illustration of the \dm profile proposed in this study.
The mass function $m(r)$ (top-left) is given by Eq.~\eqref{massF_ansatz}, with the inset highlighting the region near $r_\mathrm{ISCO}$.
The metric function $q(r)$ (top-right) is obtained from Eq.~\eqref{BmassFuncFormal} by numerical integration, where the inset  focuses on the vicinity of $r_\mathrm{ISCO}$.
The energy density $\rho(r)$ in the stationary tetrad (bottom-left), radial pressure $p_r(r)$ (bottom-right), and subsequently the metric are determined by Eqs.~\eqref{rhoDef2},~\eqref{prDef2}, and~\eqref{Mansatz}.
The calculations are carried out using the parameters $M_\mathrm{B}=1$, $a=\frac12$ (and therefore $r_\mathrm{ISCO}\simeq 4.233$), $M_\mathrm{H}=10$, $\alpha=\frac23$, $r_\mathrm{SP}=8$, and we assume $\cos\theta=\frac12$ for the calculation of the energy density and radial pressure.
We note that all the quantities manifestly satisfy the boundary conditions Eqs.~\eqref{proSce1ISO} and~\eqref{assymMF} for the mass function $m(r)$, Eqs.~\eqref{proSce1} and~\eqref{isoMF} for the metric function $q(r)$, as well as the additional physical requirements Eqs.~\eqref{prCond1} and~\eqref{prCond2} regarding the vanishing radial pressure $p_r(r)$.
}
\label{illusDKModel01}
\end{figure}

In Fig.~\ref{illusDKModel01}, we illustrate the \dm profile proposed in Eq.~\eqref{massF_ansatz}.
The calculations are carried out using the parameters $M_\mathrm{B}=1$, $a=\frac12$, $M_\mathrm{H}=10$, $\alpha=\frac23$, $r_\mathrm{SP}=8$, with the cutoff at $r_\mathrm{ISCO}\simeq 4.233$.
We note that all the quantities manifestly satisfy the boundary conditions Eqs.~\eqref{proSce1ISO} and~\eqref{assymMF} for the mass function $m(r)$, Eqs.~\eqref{proSce1} and~\eqref{isoMF} for the metric function $q(r)$, as well as the additional physical requirements Eqs.~\eqref{prCond1} and~\eqref{prCond2}, particularly the radial pressure $p_r(r)$ vanishes at spatial infinity.
Evaluating the limit in Eq.~\eqref{assLinearC} with the ISCO from Eq.~\eqref{defISCO} gives $\mathcal C\simeq19.91$ for this example.

In the static solution the normalization can be checked directly,
\bqn
4\pi\int_{r_{\rm ISCO}}^\infty r^2\rho_s(r)\,dr &=& m(\infty)-m(r_{\rm ISCO})=M_\mathrm{H}. \label{staticHaloMassIntegral}
\eqn
At finite spin $M_\mathrm{H}$ fixes the same increment in the prescribed $m(r)$, while the local matter density follows from the stress-energy tensor.

Before closing this section, we note a particular case when the integration in Eq.~\eqref{BmassFuncInv} can be carried out analytically if one assumes $\alpha=2$ for the model proposed in Eq.~\eqref{massF_ansatz}.
To perform the integration for $r \ge r_\mathrm{ISCO}$, one substitutes the explicit form of the mass function into the integrand,
\bqn
 r-2m(r) &=& r-2 M_\mathrm{B}-\frac{2 M_\mathrm{H}}{(r+r_\mathrm{SP})^2} (r-r_\mathrm{ISCO})^2 \nb\\
&=&\frac{r^3+2r^2(r_\mathrm{SP}-M_\mathrm{B}-M_\mathrm{H})+r(r_\mathrm{SP}^2-4M_\mathrm{B} r_\mathrm{SP}+4M_\mathrm{H} r_\mathrm{ISCO})-2 (M_\mathrm{B} r_\mathrm{SP}^2+M_\mathrm{H} r_\mathrm{ISCO}^2)}{(r+r_\mathrm{SP})^2} \nb\\
&=&\frac{(r-r_0)(r-r_1)(r-r_2)}{(r+r_\mathrm{SP})^2} ,\label{threeRootsDef}
\eqn
where the three roots $r_0,r_1,r_2$ are determined by
\bqn
r_0+r_1+r_2&=&-2(r_\mathrm{SP}-M_\mathrm{B}-M_\mathrm{H}) ,\nonumber \\
r_0r_1+r_0r_2+r_1r_2&=&(r_\mathrm{SP}^2-4M_\mathrm{B} r_\mathrm{SP}+4M_\mathrm{H} r_\mathrm{ISCO}) ,\nonumber\\
r_0r_1r_2&=&2 (M_\mathrm{B} r_\mathrm{SP}^2+M_\mathrm{H} r_\mathrm{ISCO}^2) .\label{rpara}
\eqn

Since all the physical parameters $M_\mathrm{B},M_\mathrm{H}, r_\mathrm{SP},r_\mathrm{ISCO}$ on the r.h.s. of the equality are positive real numbers, the three roots inevatably fall into one of three possibilities: three positive real numbers, one positive real and two negative real numbers, and a positive real and a pair of mutually conjugate complex numbers. 
Denoting $r_0$ to be the positive real root, and $r_1,r_2$ to be positive, negative, or a pair of complex numbers, we have
\bqn
\frac{1}{r-2m(r)}=\frac{\mathcal{X}}{r-r_0}+\frac{\mathcal{Y}(2r-r_1-r_2)}{2(r-r_1)(r-r_2)}+\frac{\mathcal{Z}}{(r-r_1)(r-r_2)} ,\label{polyForm}
\eqn
where
\bqn\label{cE1}
\mathcal{X}+\mathcal{Y}&=&1 ,\nonumber \\
\mathcal{X}(r_1+r_2)+\mathcal{Y}\left(\frac{r_1+r_2}{2}+r_0\right)-\mathcal{Z} &=&-2r_\mathrm{SP} ,\nonumber\\
\mathcal{X} r_1 r_2+\mathcal{Y}\frac{(r_1+r_2)r_0}{2}-\mathcal{Z} r_0&=&r_\mathrm{SP}^2 . 
\eqn
It is noted that Eqs.~\eqref{cE1} is a system of linear equations with real coefficients, and subsequently, $\mathcal{X},\mathcal{Y},\mathcal{Z}$ are manifestly real numbers. 

Using Eq.~\eqref{polyForm} and properly taking into account Eq.~\eqref{isoMF}, the integration in Eq.~\eqref{BmassFuncInv} for $r>r_\mathrm{ISCO}$ is found to be:
\bqn
&& \int\frac{dr}{2m(r)-r}=\int^r_{r_\mathrm{ISCO}}\frac{dr}{2m(r)-r}-\ln\left(r_\mathrm{ISCO}-2 M_\mathrm{B}\right)\nb \\
%&= &-\alpha \ln\left[\frac{r-r_0}{r_\mathrm{ISCO}-r_0}\right]- \frac{\beta}{2}\ln\left[\frac{r^2-(r_1+r_2)r+r_1r_2}{r^2_\mathrm{ISCO}-(r_1+r_2)r_\mathrm{ISCO}+r_1r_2}\right] \nonumber \\
%& & -\delta\left( \mathcal{I} -\mathcal{I}(r=r_\mathrm{ISCO})\right)+\ln\left(r_\mathrm{ISCO}-2 M_\mathrm{B}\right) \nb\\
& &=-\mathcal{X} \ln\left[\frac{r-r_0}{r_\mathrm{ISCO}-r_0}\right]- \frac{\mathcal{Y}}{2}\ln\left[\frac{(r-r_1)(r-r_2)}{(r_\mathrm{ISCO}-r_1)(r_\mathrm{ISCO}-r_2)}\right] -\mathcal{Z}\left(\mathcal{I}(r)-\mathcal{I}(r_\mathrm{ISCO})\right)-\ln\left(r_\mathrm{ISCO}-2 M_\mathrm{B}\right),\label{AnaSolFull}
\eqn
where
\bqn
\mathcal{I} &=&\left\{\begin{array}{ll}
    \displaystyle\frac{2}{|r_1-r_2|}\left(\arctan{\frac{2r-r_1-r_2}{|r_1-r_2|}}-\frac{\pi}{2}\right),& \mbox{if $r_1,r_2$ are complex}\\[3mm]
    \displaystyle\frac{1}{|r_1-r_2|}\ln{\frac{2r-r_1-r_2-|r_1-r_2|}{2r-r_1-r_2+|r_1-r_2|}},& \mbox{if $r_1,r_2$ are real}\
\end{array}\right.\label{defI}
\eqn

The above results can be utilized to assess the asymptotical behavior of $q(r)$ given by Eq.~\eqref{assLinearC}.
At spatial infinity, the leading contribution of Eq.~\eqref{AnaSolFull} is 
\bqn
-\mathcal{X} \ln r - \frac{\mathcal{Y}}{2} \ln r^2 =\ln r^{-(\mathcal{X}+\mathcal{Y})}=-\ln r,\label{AnaSol0}
\eqn
owing to the first line of Eq.~\eqref{cE1}, which is expected.
The constant $\mathcal{C}$ is related to the next-leading term, for which we have (cf. Eq.~\eqref{BmassFuncFormal})
\bqn
\ln \mathcal{C} = \ln\left(r_\mathrm{ISCO}-2 M_\mathrm{B}\right)-\mathcal{X} \ln\left[{r_\mathrm{ISCO}-r_0}\right] - \frac{\mathcal{Y}}{2}\ln\left[{(r_\mathrm{ISCO}-r_1)(r_\mathrm{ISCO}-r_2)}\right] -\mathcal{Z}\mathcal{I}(r_\mathrm{ISCO}),\label{AnaSol1}
\eqn
The advantage of an analytic account is that one can also derive the next-next-leading term, which turns out to furnish a vertical shift of the asymptotic straight line
\bqn
&&-\mathcal{X} \ln r \left[\frac{\ln (r-r_0)}{\ln r} -1\right] - \frac{\mathcal{Y}}{2} \ln r^2 \left[\frac{\ln (r-r_1)(r-r_2)}{\ln r^2} -1\right] -\mathcal{Z}\ \mathcal{I} \nb\\
&&\simeq -\mathcal{X} \left( -\frac{r_0}{r}\right)
- \frac{\mathcal{Y}}{2} \left(-\frac{r_1+r_2}{r}\right)
- \frac{2\mathcal{Z}}{|r_1-r_2|}\frac{(-1)|r_1-r_2|}{2r} \nb\\
&&\simeq  \left({\mathcal{X} r_0} +  \frac{\mathcal{Y}(r_1+r_2)}{2} +  {\mathcal{Z}}\right)\frac{1}{r} \equiv \frac{\mathcal{D}}{r} ,\label{AnaSol2}
\eqn
where both branches of Eq.~\eqref{defI} give $\mathcal{I}(r)=-1/r+O(r^{-2})$.
Plugging everything back into the exponential of Eq.~\eqref{BmassFuncInv}, one finds
\bqn
\frac{1}{2} \exp\left(-\int\frac{dr}{2m(r)-r}\right)
\to \frac{1}{2}\exp\left[\ln (\mathcal{C} r) -\frac{\mathcal{D}}{r}\right]
\simeq\frac{\mathcal{C} r}{2} -\frac{\mathcal{C} \mathcal{D}}{2} ,\label{anaFinalalpha2}
\eqn
where $\mathcal{C}$ and $\mathcal{D}$ are governed by Eqs.~\eqref{AnaSol1} and~\eqref{AnaSol2}.

For every $\alpha>0$, Eq.~\eqref{massF_ansatz} gives $m(r)=M_\infty+O(r^{-1})$, where $M_\infty=M_\mathrm{B}+M_\mathrm{H}$.
Eq.~\eqref{BmassFuncFormal} then gives $\partial_r\ln[(r-2q)/r]=2M_\infty/r^2+O(r^{-3})$.
Integrating from $r$ to infinity with the limit in Eq.~\eqref{assLinearC} and substituting into Eq.~\eqref{Mansatz} yields
\bqn
\frac{r-2q(r)}{r} &=& \mathcal C\left[1-\frac{2M_\infty}{r}+O(r^{-2})\right], \label{asymptoticMatchedRadial} \\
g_{TT} &=& \frac{g_{tt}}{\mathcal C}=-1+\frac{2M_\infty}{r}+O(r^{-2}) . \label{asymptoticMassLapse}
\eqn
For $\alpha=2$, comparison with Eqs.~\eqref{BmassFuncInv} and~\eqref{anaFinalalpha2} gives $\mathcal D=2M_\infty$.
The constant $\mathcal C$ evaluated in Eq.~\eqref{AnaSol1} is thus $-\lim_{r\to\infty}g_{tt}$, and its square root converts the inner Kerr time to time at infinity through $T=\sqrt{\mathcal C}\,t$.

\begin{figure}[ht]
    \centering
    \begin{minipage}{0.4\textwidth}
        \centering
        \includegraphics[width=1.0\textwidth,height=0.75\textwidth]{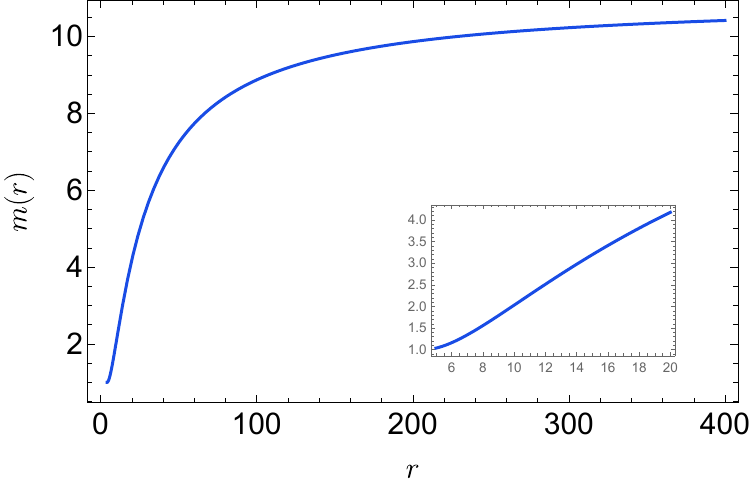}
    \end{minipage}
    \begin{minipage}{0.4\textwidth}
        \centering
        \includegraphics[width=1.0\textwidth,height=0.75\textwidth]{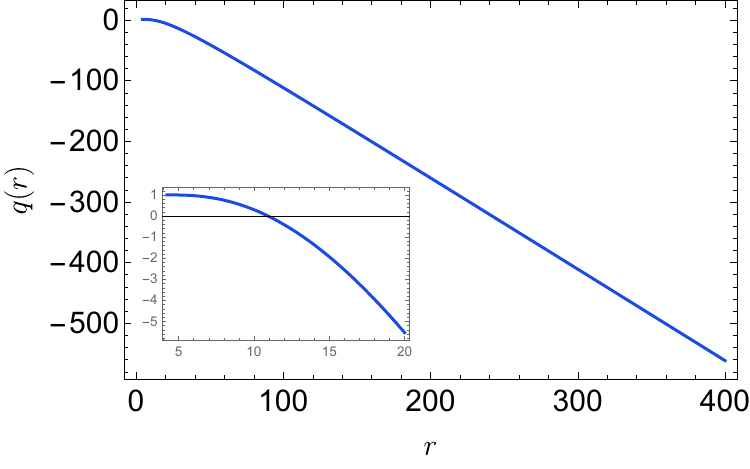}
    \end{minipage}
\renewcommand{\figurename}{Fig.}   
\caption{The assumed mass function $m(t)$ (left) and the resulting metric function $q(t)$ (right) for the \dm profile with $\alpha=2$.
The metric function are obtained using the numerical integration Eq.~\eqref{BmassFuncFormal}. 
The calculations are carried out using the parameters $M_\mathrm{B}=1$, $a=\frac12$ (and therefore $r_\mathrm{ISCO}\sim 4.25$), $M_\mathrm{H}=10$, $\alpha=2$ (NFW inner slope), and $r_\mathrm{SP}=8$.}
\label{anaDKalpha02}
\end{figure}

For an explicit evaluation of the analytic expressions, we take $M_\mathrm{B}=1$, $a=1/2$, $M_\mathrm{H}=10$, $\alpha=2$ (NFW inner slope), and $r_\mathrm{SP}=8$, and approximate the cutoff by $r_\mathrm{ISCO}=4.25$.
The three roots of Eq.~\eqref{threeRootsDef} are, numerically, $r_0=2.5321$ and $r_{1,2}= 1.73395 \pm 13.7917 i$.
This gives $\mathcal{I}(r_\mathrm{ISCO})=-0.10081$, $\mathcal{C}=4.00$, and $\mathcal{D}=22.0$.
Using the ISCO from Eq.~\eqref{defISCO} gives $\mathcal C\simeq4.024$, with the same $\mathcal D=2M_\infty$.
This can be readily compared to the results obtained via numerical integration Eq.~\eqref{BmassFuncFormal}.
From the results shown in Fig.~\ref{anaDKalpha02}, one can fit the asymptotical behavior of $q(r)$ to a linear form $Ar+B$, from which one extracts $\mathcal{C}=3.991$ and $\mathcal{D}=20.35$, in satisfactory agreement with the analytic result.

\section{Numerical results and discussions}\label{sec5}

This section elaborates on a few physically pertinent parameter sets related to empirical observations and theoretical speculations.
Recent studies have elucidated specific relationships between supermassive \bh{}s (SMBHs) and their \dm environments through distinct observational and theoretical approaches. 
Stellar orbits constrain the extended mass enclosed over the radii they probe.
For example, observations of S2 constrain the extent of a putative Galactic-centre spike for an assumed profile~\cite{agr-dark-matter-086}.
A radial slope inferred or supplied over $r_{\rm ISCO}\ll r\ll r_{\rm SP}$ can therefore be used as input for either the static or the rotating model when spin corrections are small there.
The density close to the hole can still depend on spin and polar angle.
Eq.~\eqref{rhoSpikeIntermediate} gives $\gamma_{\rm sp}=3-\alpha$ in this interval.
Cosmological simulations of cold dark matter halos give inner cusp slopes in the range $1\lesssim\gamma\lesssim1.5$, encompassing the NFW and Moore profiles~\cite{agr-dark-matter-027,agr-dark-matter-028,agr-dark-matter-042}.
For an initially cusped halo with $\rho\propto r^{-\gamma}$, the adiabatic-growth result is $\gamma_{\rm sp}=(9-2\gamma)/(4-\gamma)$~\cite{agr-dark-matter-024}.
The $\gamma=1$ case gives the $\gamma_{\rm sp}=7/3$ benchmark used in studies of the Galactic centre~\cite{agr-dark-matter-056}.
In practice, however, various dynamical processes, such as stellar heating, mergers, and \dm annihilation, can smooth out the spike and produce shallower effective slopes.
Consequently, more realistic spike profiles are expected to fall within the range $\gamma_{\rm sp} \sim 1.5$–$2.5$~\cite{agr-dark-matter-043, agr-dark-matter-044, agr-dark-matter-045}.
On this basis, we will adopt the broader interval $\gamma_{\rm sp} \sim 1$–$2.5$, which encompasses both standard halo cusps and steep \dm\ spikes, leading to
\begin{equation}
0.5 \lesssim \alpha \lesssim 2.
\end{equation}

Observations of SMBHs indicate a wide mass range $10^5 M_\odot \lesssim M_\mathrm{B} \lesssim 10^{10} M_\odot$, as inferred from stellar dynamics~\cite{agr-bh-exp-review-05} and Event Horizon Telescope (EHT) measurements~\cite{agr-strong-lensing-EHT-L01, agr-strong-lensing-EHT-L12}.
For Kerr \bh{}s, the spin parameter $a$ has the theoretical bound $0 \leq a/M_\mathrm{B} < 1$, while astrophysical observations and accretion modeling suggest that realistic systems typically satisfy $a/M_\mathrm{B} \lesssim 0.9$-$0.998$~\cite{agr-bh-exp-07, agr-bh-exp-37}.

\begin{figure}
    \centering
    \includegraphics[width=0.4\textwidth,height=0.3\textwidth]{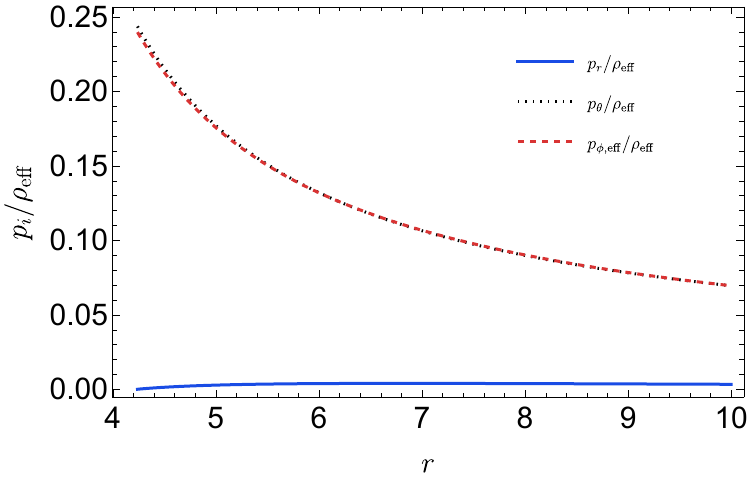}
    \caption{Radial profiles of the pressure-to-density ratios $p_{i}/\rho_{\rm eff}$ with $i=\{r,\theta,\phi_{\rm eff}\}$ for the baseline parameter set $(M_\mathrm{H},\alpha,r_{\rm SP}, a)=(10,0.5,10^4,0.5)$. 
    The principal pressures and density satisfy the energy conditions in Eq.~\eqref{energy_conditions} over the halo interval shown.
}
    \label{fig:EnergyCondition}
\end{figure}

\begin{figure}
    \centering
    \begin{minipage}{0.4\textwidth}
        \centering
        \includegraphics[width=1.0\textwidth,height=0.75\textwidth]{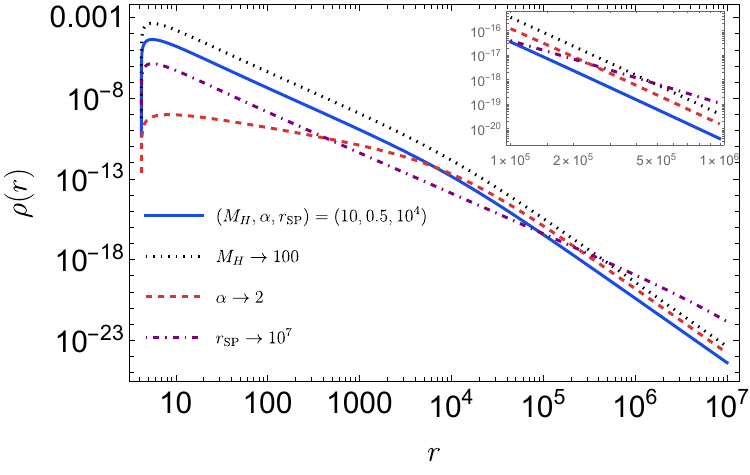}
    \end{minipage}
    \begin{minipage}{0.4\textwidth}
        \centering
        \includegraphics[width=1.0\textwidth,height=0.75\textwidth]{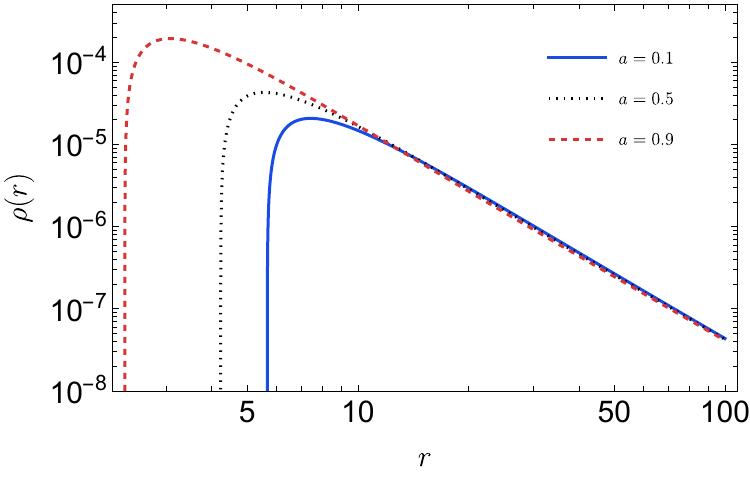}
    \end{minipage}
    \begin{minipage}{0.4\textwidth}
        \centering
        \includegraphics[width=1.0\textwidth,height=0.75\textwidth]{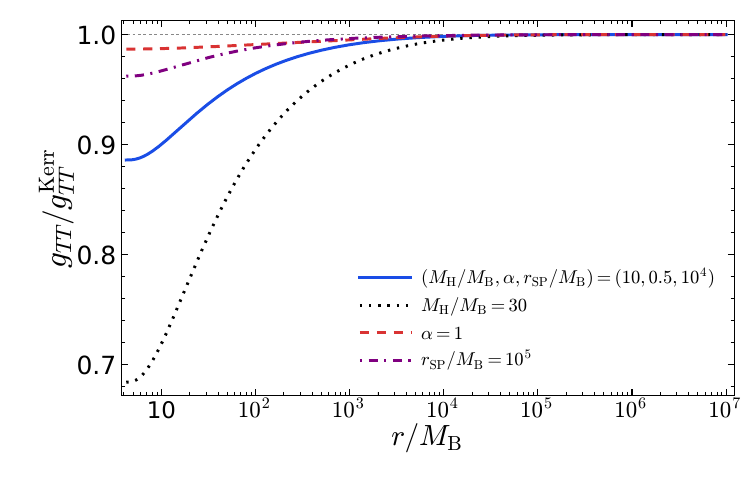}
    \end{minipage}
    \begin{minipage}{0.4\textwidth}
        \centering
        \includegraphics[width=1.0\textwidth,height=0.75\textwidth]{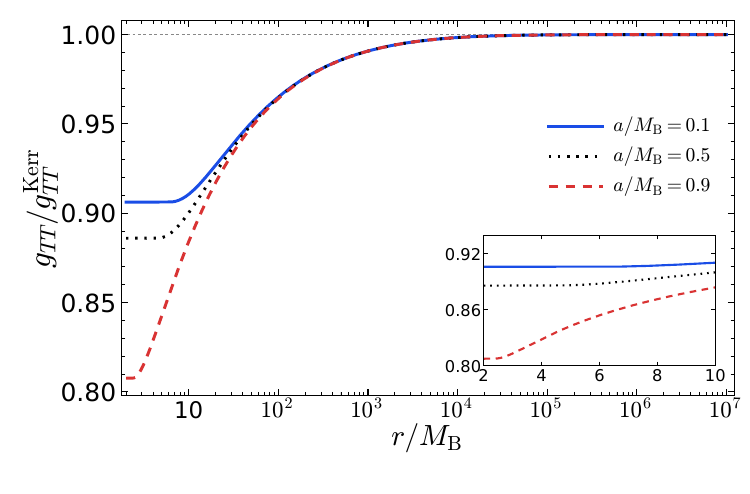}
    \end{minipage}
   \caption{The \dm energy density and the time component of the metric for the proposed model. 
   Unless specified otherwise, the calculations use the baseline parameter set $(M_\mathrm{H},\alpha,r_{\rm SP}, a)=(10,0.5,10^4,0.5)$. 
   The top panels show the radial density $\rho(r)$ measured in the stationary tetrad. 
   The bottom panels show $g_{TT}/g_{TT}^{\rm Kerr}$ at $\cos\theta=1/2$, with time normalized at infinity for both metrics.
   The horizontal line at unity is the common asymptote of all the curves.
   The left panels vary one halo parameter at a time about the baseline values.
   The right panels vary the spin parameter $a$.
   The inset in the lower right panel shows where each curve departs from the inner value $1/\mathcal C$.}
    \label{fig:Density}
\end{figure}

In the static model, the halo mass enclosed within a finite radius is $m(r)-M_\mathrm{B}$, a fraction of $M_\mathrm{H}$.
The amount of dark matter in a galactic centre depends on the profile and the radius probed. 
For the Milky Way, the \bh mass is $M_\mathrm{B} \simeq 4\times10^6 M_\odot$, while the \dm mass within the central parsec can reach $\sim 10^7$-$10^8 M_\odot$, depending on the assumed density profile~\cite{agr-dark-matter-024, agr-dark-matter-043}. 
For M87*, with $M_\mathrm{B} \simeq 6.5\times10^9 M_\odot$, the \dm concentration in the central region can also be substantial~\cite{agr-bh-exp-38}. 
These estimates motivate considering halo masses in the range $M_\mathrm{H} \sim 1$-$10^3\,M_\mathrm{B}$. 
Theoretical models of \dm spikes around \bh{}s indicate that the spike extends over sub-parsec to parsec scales. 
For the Galactic center one finds $r_{\rm SP} \sim 0.1$-$1\,\mathrm{pc}$, while for more massive systems such as M87* it can reach $\sim 10\,\mathrm{pc}$~\cite{agr-dark-matter-024, agr-dark-matter-059}. 
This corresponds to $r_{\rm SP} \sim 10^4$-$10^7 M_\mathrm{B}$ and implies a hierarchy $r_{\rm SP} \gg M_\mathrm{H} \gg r_{\rm ISCO}$.
Putting all the pieces together, we adopt the parameter ranges summarized in Tab.~\ref{tab:parameters}, and the corresponding numerical results are presented in Fig.~\ref{fig:EnergyCondition} and Fig.~\ref{fig:Density}.

\begin{table}[th]
\centering
\begin{tabular}{c c}
\hline\hline
Parameter & Physical range \\
\hline
$M_\mathrm{B}$ & $10^5$-$10^{10}\,M_\odot$ \\
$a/M_\mathrm{B}$ & $0$-$0.998$ \\
$\alpha$ & $0.5$-$2$ \\
$M_\mathrm{H}/M_\mathrm{B}$ & $1$-$10^{3}$ \\
$r_{\rm SP}/M_\mathrm{B}$ & $10^4$-$10^7$ \\
\hline\hline
\end{tabular}
\caption{
Astrophysically motivated parameter ranges for supermassive black holes and their surrounding dark-matter halos, inferred from observations and theoretical models of halo cusps and spikes.
}
\label{tab:parameters}
\end{table}

In addition to the constraints inferred from empirical observations, these parameters must also satisfy physical requirements such as subluminality and the energy conditions.
To this end, for each parameter combination, we first examine the conditions given by Eq.~\eqref{condTtphi}.
When these conditions hold, the stress-energy tensor admits a local rest frame and can be diagonalized by a Lorentz boost in the $(\hat t,\hat\phi)$ plane (see Appx.~\ref{appB}).
The eigenvalues of the mixed $(\hat t,\hat\phi)$ block are $-\rho_{\rm eff}$ and $p_{\phi,\rm eff}$, where $\rho_{\rm eff}$ is the matter rest-frame density given in Eq.~\eqref{rhoPrincipal}.
The energy conditions~\cite{agr-bh-exp-39,book-cosmology-Hawking} are then evaluated in terms of the principal energy density and pressures as
\begin{equation}
\begin{aligned}
& \mathrm{SEC}: &&
\rho_{\rm eff}+\sum_i p_i\geq0,
\qquad
\rho_{\rm eff}+p_i\geq0,\\
& \mathrm{WEC}: &&
\rho_{\rm eff}\geq0,
\qquad
\rho_{\rm eff}+p_i\geq0,\\
& \mathrm{NEC}: &&
\rho_{\rm eff}+p_i\geq0,\\
& \mathrm{DEC}: &&
\rho_{\rm eff}\geq |p_i|,
\end{aligned}
\label{energy_conditions}
\end{equation}
where $i\in\{r,\theta,\phi_{\rm eff}\}$ and $p_{\phi,\rm eff}$ denotes the principal azimuthal pressure.

At the radii sampled in the inner-halo survey, no violation of the WEC or SEC was found for any of the sampled configurations with $a/M_B\leq0.8$.
For $a/M_B>0.8$, violations of these conditions may occur for some extreme combinations of the halo parameters, particularly those involving a large halo mass, a small spike radius, and a steep inner density profile.
Therefore, $a/M_B\simeq0.8$ may be regarded as an empirical boundary of the present numerical survey rather than a universal analytic limit.
As an illustration, Fig.~\ref{fig:EnergyCondition} presents the radial profiles of the pressure-to-density ratios, $p_i/\rho_{\rm eff}$, for the baseline model characterized by $M_\mathrm{H}=10M_\mathrm{B}$, $\alpha=0.5$, $r_{\rm SP}=10^4M_\mathrm{B}$, and $a=0.5$.
For this configuration, the SEC, WEC, NEC, and DEC are satisfied throughout the radial interval shown in the figure.
All dimensional quantities shown in the figure are normalized by $M_\mathrm{B}$.
On the equator, the smaller of the two contractions with the azimuthal null vectors $e_{\hat t}\pm e_{\hat\phi}$ is $\rho+p_\phi-2|T_{\hat t\hat\phi}|$.
For the baseline model this quantity changes from positive to negative at $r\simeq8.12\times10^5 M_\mathrm{B}$, where the timelike rest-frame condition in Eq.~\eqref{condTtphi} reaches equality and the null energy condition fails just outside that radius.
For the compact halo of Fig.~\ref{illusDKModel01}, the corresponding equatorial crossing occurs at $r\simeq69.1M_\mathrm{B}$.

We now turn to the main features of the metric modification induced by the \dm distribution, as illustrated in Fig.~\ref{fig:Density}. 
As one observes in the top-left panel, the radial density profile is controlled by the three \dm parameters $(M_\mathrm{H},\alpha,r_{\mathrm{SP}})$, each with a distinct physical role. 
The halo mass $M_\mathrm{H}$ primarily rescales the overall normalization of the density while leaving the profile shape largely unchanged. 
In the interval $r_{\rm ISCO}\ll r\ll r_{\rm SP}$, increasing $\alpha$ makes the profile shallower, as in Eq.~\eqref{rhoSpikeIntermediate}.
The scale $r_{\rm SP}$ sets the transition to the outer profile.

The \bh spin introduces a qualitatively distinct effect. 
As shown in the top-right panel of Fig.~\ref{fig:Density}, varying the spin parameter $a$ primarily shifts the location of the \isco, which acts as an effective inner cutoff in the present model. 
Decreasing $a$ moves the \isco to larger radii, truncating the density profile further out and shifting the peak accordingly, while increasing $a$ allows the distribution to extend closer to the \bh, resulting in a higher and more centrally concentrated peak. 
Beyond the peak region, the profiles rapidly converge, indicating that the large-scale halo structure is essentially insensitive to the spin.

To compare the rotating matter density with the static prediction, we evaluate the principal density $\rho_{\rm eff}$ from Eq.~\eqref{rhoPrincipal}.
For the baseline of Fig.~\ref{fig:Density} at $\cos\theta=1/2$, the ratio $\rho_{\rm eff}(a=0.5M_\mathrm{B})/\rho_s(a=0)$ is $1.167$, $0.9710$, and $0.9956$ at $r/M_\mathrm{B}=10$, $100$, and $1000$, respectively.
Each model uses its own ISCO.
At the latter two radii the transformation from the stationary tetrad to the matter rest frame changes the rotating density by $0.19\%$ and $0.034\%$, respectively.
To test the power law well inside the spike, we increase only $r_{\rm SP}$ to $10^7 M_\mathrm{B}$.
Over $10^3\le r/M_\mathrm{B}\le10^4$, the logarithmic slope $-\Delta\ln\rho/\Delta\ln r$ is $2.505$ for the static density and $2.504$ for the rotating principal density at $\cos\theta=0$, $1/2$, and $1$.
Both agree to within $0.006$ with the intermediate exponent $3-\alpha=2.5$.

The bottom panels of Fig.~\ref{fig:Density} compare the halo metric with Kerr at the same $M_\mathrm{B}$ and $a$, with time normalized at spatial infinity in both spacetimes.
In the matching coordinate $t$, the ratio $g_{tt}/g_{tt}^{\rm Kerr}$ is unity in the inner vacuum wherever it is defined and tends to $\mathcal C$ at infinity.
Since the usual Kerr time already has the chosen normalization, the plotted ratio is $g_{TT}/g_{TT}^{\rm Kerr}=(g_{tt}/g_{tt}^{\rm Kerr})/\mathcal C$ and its inner value is $1/\mathcal C$.
For the baseline parameters, Eq.~\eqref{assLinearC} gives $\mathcal C=1.12867$ and hence an inner ratio of $0.88600$.
At a fixed position in the inner vacuum where $\partial_T$ is timelike, a stationary clock therefore runs at $1/\sqrt{\mathcal C}=0.94128$ times the rate at the same position in isolated Kerr when both rates are referred to infinity.

In the lower left panel of Fig.~\ref{fig:Density}, increasing $M_\mathrm{H}/M_\mathrm{B}$ from $10$ to $30$ lowers the ratio at the inner edge from $0.886$ to $0.684$.
At fixed $M_\mathrm{H}$, increasing $\alpha$ or $r_{\rm SP}$ reduces the mass enclosed near the black hole in Eq.~\eqref{massF_ansatz} and brings the inner ratio closer to unity.
In the lower right panel the halo parameters are fixed, and the spin affects the inner redshift through $r_{\rm ISCO}$ in the mass profile and the matching condition.
Increasing $a/M_\mathrm{B}$ from $0.1$ to $0.9$ moves the prograde ISCO from $5.67M_\mathrm{B}$ to $2.32M_\mathrm{B}$ and lowers the inner ratio from $0.906$ to $0.808$.
Each curve rises outside its own cutoff and approaches unity from below.
At large radius the ratio is $1-2M_\mathrm{H}/r+O(r^{-2})$, so all curves with the same halo mass share the same leading departure from unity, irrespective of the spin or the shape of the inner spike.

\section{Further discussions and concluding remarks}\label{sec6}

In this work, we constructed an analytic metric describing rotating black holes surrounded by a generic dark-matter halo with a central spike, providing an exact solution of the Einstein field equations. 
The model consistently incorporates a physically motivated truncation of the dark-matter distribution near the horizon, ensuring that both the energy density and radial pressure vanish in the near-horizon region. 
The construction yields an anisotropic stress-energy tensor and recovers the Kerr solution in the absence of the halo.
Conservation of orbital actions during adiabatic black-hole growth predicts a spike in an initially cusped collisionless halo, including when the final hole rotates~\cite{agr-dark-matter-024,agr-dark-matter-059,agr-dark-matter-087}.
Following Cardoso {\it et al.}~\cite{agr-BH-spectroscopy-024}, we take the radial mass profile as input to the field equations.
Eq.~\eqref{massF_ansatz} gives an analytic example with a tunable spike slope and allows the integral for $q(r)$ to be evaluated in closed form when $\alpha=2$.
Its parameters specify the halo without solving for the particle distribution during growth.
A tabulated radial mass profile from a dynamical calculation can be used in the same construction.
If the input is a spherical density, Eq.~\eqref{asymRho} first gives the enclosed mass.
For any such profile satisfying the inner matching and regularity conditions, numerical quadrature of Eq.~\eqref{BmassFuncFormal} then gives $q(r)$.

A relativistic halo model must specify how the matter is supported against gravity.
The Einstein cluster describes collisionless particles on circular orbits, giving zero radial pressure and a tangential pressure fixed by conservation, as in the spherical halo solutions of Cardoso {\it et al.} and Figueiredo {\it et al.}~\cite{agr-BH-spectroscopy-024,agr-dark-matter-070}.
Nampalliwar {\it et al.}~\cite{agr-dark-matter-098} used a power-law spike outside an inner radius $r_{\rm b}$ to study S2 and the shadow of Sgr~A*, exploring two choices of isotropic pressure and introducing rotation with the Newman-Janis-Azreg-A\"{\i}nou construction.
In spherical symmetry the density fixes the enclosed mass, and the radial pressure contributes separately to the lapse equation.
For an isotropic equation of state, the pressure gradient associated with the chosen density must also satisfy hydrostatic balance~\cite{agr-dark-matter-093}.
Finite radial velocity dispersion can generate radial pressure even in a collisionless halo.
Zhao and Pani~\cite{agr-BH-spectroscopy-068} generalized the spherical Einstein cluster by prescribing a polytropic law $p_r=\mathcal K\rho^\gamma$ and determining $p_t$ from the anisotropic hydrostatic equation.
The limit $\mathcal K=0$ recovers the Einstein cluster for the same mass profile, and the dilute-halo limit connects to expansions about a vacuum black hole.
For their extended Hernquist-type model the leading shifts of circular-orbit frequencies and axial quasinormal modes are governed by the halo redshift, which is strengthened by radial pressure~\cite{agr-BH-spectroscopy-068}.
Magnetic tidal Love numbers are more sensitive to pressure anisotropy and to whether the fluid response is taken to be strictly static or irrotational~\cite{agr-BH-spectroscopy-067,agr-BH-spectroscopy-068}.
Fernandes and Cardoso~\cite{agr-BH-spectroscopy-062} obtained stationary, axisymmetric solutions numerically while imposing vanishing radial pressure in the rotating halo.
Here we adopt the Einstein-cluster prescription in the spherical seed, where $p_r=0$ and conservation gives $p_t$ in Eq.~\eqref{asymPt} without a separate equation of state.
Our mass profile gives the intermediate density law in Eq.~\eqref{rhoSpikeIntermediate} and permits smooth matching to a Kerr interior at the prograde ISCO.
The rotating metric in Eq.~\eqref{Mansatz} recovers the spherical seed exactly at zero spin.
At finite spin the field equations give the halo stresses and azimuthal momentum density, with $p_r$ vanishing in the Kerr interior and tending to zero at infinity as specified in Eqs.~\eqref{prCond1} and~\eqref{prCond2}.

Physically relevant parameter ranges are also discussed, following the observations of SMBHs and theoretical models of dark-matter halos.
One considers, inclusively, the scenario where the halo mass is comparable to or significantly exceeds the black-hole mass, while the spike typically extends over several orders of magnitude in radius. 
For the baseline parameters, the energy conditions hold over the interval shown in Fig.~\ref{fig:EnergyCondition}, while the equatorial null energy condition fails sufficiently far out in the halo.
For the profiles in Fig.~\ref{fig:Density}, a more massive or more centrally concentrated halo produces a larger inner redshift, and increasing the prograde spin moves the inner edge inward and strengthens that redshift.
With time normalized at infinity, the total mass $M_\mathrm{B}+M_\mathrm{H}$ fixes the $1/r$ term in $g_{TT}$, while the inner clock rate relative to infinity depends on the halo profile through $\mathcal C$.

The resulting metric provides an analytically tractable setting for investigating environmental effects on black-hole spacetimes. 
Further studies will address the stability of the resulting spacetime and observational constraints from current and forthcoming high-precision measurements. 
The obtained metric is directly applicable to a broad range of astrophysical phenomena, including gravitational lensing, black-hole shadows, and gravitational-wave signals from compact-object inspirals. 
We plan to pursue some of these directions in subsequent studies.

\section*{Acknowledgements}

This work is supported by the National Key Research and Development Program of China under Grant No. 2020YFC2201400.
We also gratefully acknowledge the financial support from Brazilian agencies Funda\c{c}\~ao de Amparo \`a Pesquisa do Estado de S\~ao Paulo (FAPESP), Funda\c{c}\~ao de Amparo \`a Pesquisa do Estado do Rio de Janeiro (FAPERJ), Conselho Nacional de Desenvolvimento Cient\'{\i}fico e Tecnol\'ogico (CNPq), and Coordena\c{c}\~ao de Aperfei\c{c}oamento de Pessoal de N\'ivel Superior (CAPES).
A part of this work was developed under the project Institutos Nacionais de Ci\^{e}ncias e Tecnologia - F\'isica Nuclear e Aplica\c{c}\~{o}es (INCT/FNA) Proc. No. 408419/2024-5.
This research is also supported by the Center for Scientific Computing (NCC/GridUNESP) of S\~ao Paulo State University (UNESP).
YZ is supported by the MUR FIS2 Advanced Grant ET-NOW (CUP:~B53C25001080001), the INFN TEONGRAV initiative, and the Chinese Scholarship Council (CSC).

\appendix

\section{Local rest frame of the dark matter}\label{appB}

The orthonormal tetrad introduced in Sec.~\ref{sec2} corresponds to a stationary observer with respect to the background rotating black-hole spacetime.
Since the matter distribution is itself rotating, the corresponding stress-energy tensor Eq.~\eqref{Tab} possesses a nonvanishing momentum density $T_{\hat t\hat\phi}$.
The principal energy density and principal pressures are instead defined in the local rest frame of the matter, in which the momentum density vanishes.
Such a frame is related to the stationary tetrad by a Lorentz boost in the $(\hat t,\hat\phi)$ plane,
\begin{equation}
\Lambda^{\hat a}{}_{\hat b}
=
\begin{pmatrix}
\gamma & 0 & 0 & -\gamma v\\
0&1&0&0\\
0&0&1&0\\
-\gamma v&0&0&\gamma
\end{pmatrix},
\qquad
\gamma=\frac{1}{\sqrt{1-v^2}},
\end{equation}
where $v$ denotes the azimuthal velocity of the matter relative to the stationary observer.

The transformed stress-energy tensor is
\begin{equation}
T'_{\hat a\hat b}
=
\Lambda^{\hat c}{}_{\hat a}
\Lambda^{\hat d}{}_{\hat b}
T_{\hat c\hat d},
\end{equation}
whose off-diagonal component is
\begin{equation}
T_{\hat t'\hat\phi'}
=
\gamma^2
\left[
T_{\hat t\hat\phi}(1+v^2)-v(\rho+p_\phi)
\right].
\end{equation}
The local rest frame is obtained by requiring
\begin{equation}
T_{\hat t'\hat\phi'}=0,
\end{equation}
which leads to the quadratic equation
\begin{equation}
T_{\hat t\hat\phi}v^2-(\rho+p_\phi)v+T_{\hat t\hat\phi}=0.
\end{equation}

For a nonzero momentum density, the branch with positive $\rho+p_\phi$ admits a timelike rest frame when
\begin{equation}
(\rho+p_\phi)^2>4T_{\hat t\hat\phi}^{\,2}.
\end{equation}
Equality would give $|v|=1$.
If $T_{\hat t\hat\phi}=0$, the stationary tetrad already diagonalizes the stress-energy tensor.

The corresponding boost velocity is
\begin{equation}
v
=
\frac{(\rho+p_\phi)
-
\sqrt{(\rho+p_\phi)^2-4T_{\hat t\hat\phi}^2}}
{2T_{\hat t\hat\phi}},
\label{boostvelocity}
\end{equation}
where the branch has been chosen so that $|v|<1$ and $v\rightarrow0$ as $T_{\hat t\hat\phi}\rightarrow0$.

In the local rest frame the stress-energy tensor becomes diagonal,
\begin{equation}
T'_{\hat a\hat b}
=
\mathrm{diag}
\left(
\rho_{\rm eff},
p_r,
p_\theta,
p_{\phi,\rm eff}
\right),
\end{equation}
where diagonalization of the $(\hat t,\hat\phi)$ block gives
\bqn
\rho_{\rm eff} &=& \frac12\left[\rho-p_\phi+\sqrt{(\rho+p_\phi)^2-4T_{\hat t\hat\phi}^2}\right], \label{rhoPrincipal} \\
p_{\phi,\rm eff} &=& \frac12\left[-\rho+p_\phi+\sqrt{(\rho+p_\phi)^2-4T_{\hat t\hat\phi}^2}\right]. \label{azimuthalPrincipalPressure}
\eqn
The above conditions are equivalent to requiring that the stress-energy tensor belongs to the Hawking-Ellis Type-I class~\cite{book-cosmology-Hawking}.

\section{Comparison with Newman-Janis construction and its alternatives}\label{appA}

In this Appendix, we compare the metric derived in the present study against the Newman-Janis algorithm~\cite{agr-Newman-Janis-01, agr-Newman-Janis-02} and its alternatives~\cite{agr-bh-Kerr-55}.

Specifically, we aim to establish a relation between the rotating metric proposed in Eq.~\eqref{Mansatz} and that obtained using the Newman-Janis algorithm.
The metric presented in the main text was derived independently by solving Einstein's field equations, rather than by applying the Newman-Janis prescription.
Nonetheless, our purpose here is not to derive the rotating solution from the Newman-Janis algorithm, but to investigate this possibility by examining whether its associated Newman-Penrose null tetrad is algebraically equivalent to the null tetrad obtained from the Newman-Janis transformation of the corresponding static seed metric, provided that an appropriate complexification scheme is adopted.
We show that no such complexification scheme can be established.
This leads us to conclude that the proposed metric differs conceptually from those obtained using the Newman-Janis algorithm and its known alternatives.

Throughout this Appendix, we follow the Newman-Penrose~\cite{book-blackhole-Chandrasekhar} formalism where the null tetrad satisfies
\begin{equation}
g^{\mu\nu}
=
-l^\mu n^\nu
-l^\nu n^\mu
+m^\mu\bar m^\nu
+\bar m^\mu m^\nu,
\label{eq:NPmetric}
\end{equation}
together with the orthogonality conditions
\begin{equation}
l^\mu l_\mu
=
n^\mu n_\mu
=
m^\mu m_\mu
=
0,
\end{equation}
\begin{equation}
l^\mu n_\mu=-1,
\qquad
m^\mu\bar m_\mu=1,
\end{equation}
with all remaining inner products being zero.

\subsection{The static seed metric}

For the null-tetrad calculation, we use the static lapse function $f_\ast(r)$ defined in Sec.~\ref{sec2} and write Eq.~\eqref{MSansatz} as
\begin{equation}
ds^2
=
-f_\ast(r)\,dt^2
+
\frac{dr^2}{G(r)}
+
r^2
\left(
d\theta^2
+
\sin^2\theta\,d\phi^2
\right),
\label{eq:Astaticmetric}
\end{equation}
where
\begin{equation}
G(r)
=
1-\frac{2m(r)}{r} .
\label{eq:Gdef}
\end{equation}

The Newman-Janis algorithm is most conveniently carried out in advanced Eddington-Finkelstein coordinates. 
Introducing the null coordinate
\begin{equation}
du
=
dt
-
\frac{dr}
{\sqrt{f_\ast(r)G(r)}},
\label{eq:EFtransform}
\end{equation}
the metric becomes
\begin{equation}
ds^2
=
-f_\ast(r)\,du^2
-
2A(r)\,du\,dr
+
r^2d\Omega^2,
\label{eq:EFmetric}
\end{equation}
where
\begin{equation}
A(r)
=
\sqrt{\frac{f_\ast(r)}{G(r)}}.
\label{eq:Adef}
\end{equation}

The corresponding null tetrad is readily obtained as
\bqn\label{eq:lnmSeed}
l^\mu
&=&
\delta_r^\mu,\nb\\
n^\mu
&=&
\frac1{A(r)}
\delta_u^\mu
-
\frac{G(r)}2
\delta_r^\mu,\nb\\
m^\mu
&=&
\frac1{\sqrt2\,r}
\left(
\delta_\theta^\mu
+
\frac{i}{\sin\theta}
\delta_\phi^\mu
\right),\nb\\
\bar m^\mu
&=&
\frac1{\sqrt2\,r}
\left(
\delta_\theta^\mu
-
\frac{i}{\sin\theta}
\delta_\phi^\mu
\right).
\eqn

The identifications in Sec.~\ref{sec2} give
\begin{equation}
f(r)
=
r^2
\left[
1-f_\ast(r)
\right],
\label{eq:Fdef}
\end{equation}
and
\begin{equation}
B(r)
=
2m(r)r.
\tag{\ref{massFunc}}
\end{equation}
Subsequently, the metric function Eq.~\eqref{eq:Adef} can be expressed as
\begin{equation}
A^2(r)
=
\frac
{1-\dfrac{f(r)}{r^2}}
{1-\dfrac{B(r)}{r^2}}.
\label{eq:AFB}
\end{equation}

\subsection{Newman-Janis transformation of the null tetrad}

The original Newman-Janis algorithm proceeds by allowing the coordinates $u,\;r$ to take complex values while preserving the Newman-Penrose normalization conditions. 
The null basis vectors are then regarded as functions of the complex variables $(u,r,\bar r)$, where a specific complex coordinate transformation is subsequently adopted.

The standard Newman-Janis transformation is
\begin{equation}
u'
=
u-ia\cos\theta,
\qquad
r'
=
r+ia\cos\theta,
\label{eq:NJcoord}
\end{equation}
where $a$ is the rotation parameter.

Under Eq.~\eqref{eq:NJcoord}, the Jacobian gives
\begin{equation}
\frac{\partial}{\partial r}
=
\frac{\partial}{\partial r'},
\end{equation}
and
\begin{equation}
\frac{\partial}{\partial\theta}
=
\frac{\partial}{\partial\theta'}
+
ia\sin\theta
\left(
\frac{\partial}{\partial u'}
-
\frac{\partial}{\partial r'}
\right).
\end{equation}
Suppressing the primes for simplicity, the transformed null tetrad becomes
\bqn\label{eq:lnmrot}
l^\mu 
&=&
\delta_r^\mu ,\nb\\
n^\mu
&=&
\frac{1}{\widetilde A}
\delta_u^\mu
-
\frac{\widetilde G}{2}
\delta_r^\mu ,\nb\\
m^\mu
&=&
\frac{1}
{\sqrt2\left(r+ia\cos\theta\right)}
\left[
ia\sin\theta
\left(
\delta_u^\mu
-
\delta_r^\mu
\right)
+
\delta_\theta^\mu
+
\frac{i}{\sin\theta}
\delta_\phi^\mu
\right],\nb\\
\bar m^\mu
&=&
\frac{1}
{\sqrt2\left(r-ia\cos\theta\right)}
\left[
-ia\sin\theta
\left(
\delta_u^\mu
-
\delta_r^\mu
\right)
+
\delta_\theta^\mu
-
\frac{i}{\sin\theta}
\delta_\phi^\mu
\right].
\eqn
where the complexification relations $G\to {\widetilde G}$ and $A\to {\widetilde A}$ are to be determined.

Noting, $\Sigma = \left(r+ia\cos\theta\right)\left(r-ia\cos\theta\right)$, the angular part satisfies
\begin{equation}
m^\mu\bar m^\nu+\bar m^\mu m^\nu
=
\frac1\Sigma
\begin{pmatrix}
a^2\sin^2\theta
&
-a^2\sin^2\theta
&
0
&
a
\\
-a^2\sin^2\theta
&
a^2\sin^2\theta
&
0
&
-a
\\
0
&
0
&
1
&
0
\\
a
&
-a
&
0
&
\dfrac1{\sin^2\theta}
\end{pmatrix} .
\label{eq:mmbar}
\end{equation}
Similarly,
\begin{equation}
-l^\mu n^\nu
-l^\nu n^\mu
=
\begin{pmatrix}
0
&
-\dfrac1{\widetilde A}
&
0
&
0
\\
-\dfrac1{\widetilde A}
&
\widetilde G
&
0
&
0
\\
0
&
0
&
0
&
0
\\
0
&
0
&
0
&
0
\end{pmatrix}.
\label{eq:lnmatrix}
\end{equation}

Adding Eqs.~\eqref{eq:mmbar} and~\eqref{eq:lnmatrix} gives the reconstructed inverse metric,
\begin{equation}
g^{\mu\nu}
=
\begin{pmatrix}
\dfrac{a^2\sin^2\theta}{\Sigma}
&
-\dfrac1{\widetilde A}
-
\dfrac{a^2\sin^2\theta}{\Sigma}
&
0
&
\dfrac a\Sigma
\\
-\dfrac1{\widetilde A}
-
\dfrac{a^2\sin^2\theta}{\Sigma}
&
\widetilde G
+
\dfrac{a^2\sin^2\theta}{\Sigma}
&
0
&
-\dfrac a\Sigma
\\
0
&
0
&
\dfrac1\Sigma
&
0
\\
\dfrac a\Sigma
&
-\dfrac a\Sigma
&
0
&
\dfrac1{\Sigma\sin^2\theta}
\end{pmatrix}.
\label{eq:ginvrot}
\end{equation}

Up to this point, no assumption has been made regarding the explicit form of the complexified functions $\widetilde A$ and $\widetilde G$. 
They must be defined for individual terms of the Taylor expansion of respective metric functions, and are subject to specific choice.
Otherwise, Eq.~\eqref{eq:ginvrot} is a general result to all Newman-Janis constructions based on the static metric in Eq.~\eqref{eq:Astaticmetric}.
The only remaining freedom lies in the prescription adopted for the complexification of the radial functions.

\subsection{Generalized Eddington-Finkelstein coordinates of the rotating metric}

We now derive the generalized Eddington-Finkelstein (EF) coordinates directly from the rotating metric Eq.~\eqref{Mansatz}.
Unlike the Kerr spacetime, where the Boyer-Lindquist to EF transformation is known a priori, here the metric contains two independent radial functions $f(r)$ and $B(r)$.
Therefore, the coordinate transformation must be determined from the metric itself rather than assumed from the Kerr solution.

We consider the general coordinate transformation
\begin{equation}
dt
=
du+\lambda(r)\,dr,
\qquad
d\phi
=
d\varphi+\chi(r)\,dr,
\label{eq:EFgeneral}
\end{equation}
where the functions $\lambda(r)$ and $\chi(r)$ remain to be determined.

Substituting Eq.~\eqref{eq:EFgeneral} into Eq.~\eqref{Mansatz} gives
\begin{equation}
\begin{aligned}
ds^2
=&
g_{tt}du^2
+
2g^{(\rm EF)}_{ur}du\,dr
+
2g^{(\rm EF)}_{r\varphi} dr\,d\varphi
+
g^{(\rm EF)}_{rr} dr^2
+
2g_{t\phi}du\,d\varphi
+
g_{\phi\phi}d\varphi^2
+
g_{\theta\theta}d\theta^2 .
\end{aligned}
\label{eq:EFexpanded}
\end{equation}
where the transformed metric components are
\begin{equation}
g^{(\rm EF)}_{ur}
=
g_{tt}\lambda
+
g_{t\phi}\chi ,
\label{eq:gur}
\end{equation}
\begin{equation}
g^{(\rm EF)}_{r\varphi}
=
g_{t\phi}\lambda
+
g_{\phi\phi}\chi ,
\label{eq:grphi}
\end{equation}
and
\begin{equation}
g^{(\rm EF)}_{rr}
=
g_{rr}
+
g_{tt}\lambda^2
+
2g_{t\phi}\lambda\chi
+
g_{\phi\phi}\chi^2 .
\label{eq:grrEF}
\end{equation}

The transformed spacetime is required to satisfy two conditions.
First, the radial coordinate should become null,
\begin{equation}
g^{(\rm EF)}_{rr}=0.
\label{eq:nullcondition}
\end{equation}
Second, the transformation is required to possess the Kerr-Schild structure, namely, that the two functions are proportional with the same ratio as in the Kerr spacetime,
\begin{equation}
\frac{\chi}{\lambda} =\frac{a}{r^2+a^2} .
\label{eq:KScondition}
\end{equation}
Together with Eqs.~\eqref{eq:grphi} and~\eqref{eq:grrEF}, the conditions in Eqs.~\eqref{eq:nullcondition} and~\eqref{eq:KScondition} determine the unknown functions $\lambda(r)$ and $\chi(r)$ up to an overall sign.

Before solving Eqs.~\eqref{eq:nullcondition} and~\eqref{eq:KScondition}, we first establish an algebraic identity satisfied by the metric coefficients of Eq.~\eqref{Mansatz}:
\begin{equation}
g_{tt}g_{\phi\phi}
-
g_{t\phi}^{\,2}
=
-
\left(
r^2+a^2-f(r)
\right)
\sin^2\theta ,
\label{eq:blockidentity}
\end{equation}
which is obtained straightforwardly by using the explicit metric coefficients.
It is noted that the Kerr identity is recovered in the vacuum case, where $f(r)=B(r)=2Mr$.

We now proceed to solve Eqs.~\eqref{eq:KScondition} and~\eqref{eq:nullcondition}.
Instead of substituting Eq.~\eqref{eq:KScondition} directly into Eq.~\eqref{eq:nullcondition}, it is convenient to introduce an auxiliary radial function $\beta(r)$ through
\begin{equation}
\lambda
=
\frac{r^2+a^2}
{r^2+a^2-B}
\,\beta(r) ,
\label{eq:gdef}
\end{equation}
and
\begin{equation}
\chi
=
\frac{a}
{r^2+a^2-B}
\,\beta(r),
\label{eq:chi2}
\end{equation}
which manifestly satisfy the Kerr-Schild condition in Eq.~\eqref{eq:KScondition} for any $\beta(r)$.

Substituting Eqs.~\eqref{eq:gdef} and~\eqref{eq:chi2} into Eq.~\eqref{eq:nullcondition}, and employing the identity in Eq.~\eqref{eq:blockidentity}, yields
\begin{equation}
\beta^2(r)
\,
\frac
{r^2+a^2-f}
{r^2+a^2-B}
=
1.
\label{eq:betaeq}
\end{equation}
In other words, 
\begin{equation}
\beta(r)
=
\varepsilon
\sqrt{
\frac
{r^2+a^2-B}
{r^2+a^2-f}
} ,
\qquad
\varepsilon=\pm 1 ,
\label{eq:gsolution}
\end{equation}
where the residual sign $\varepsilon$ is not fixed by Eqs.~\eqref{eq:nullcondition} and~\eqref{eq:KScondition}, as both are homogeneous in $\beta$.
We take $\varepsilon=+1$, which is the branch consistent with the convention adopted for the static seed metric, Eq.~\eqref{eq:Astaticmetric}; equivalently, it is the branch for which the standard EF form of the Kerr metric is recovered when $f(r)=B(r)$, as verified below.

Finally, we have
\begin{equation}
\lambda
=
\frac{r^2+a^2}
{\sqrt{(r^2+a^2-B)(r^2+a^2-f)}},
\label{eq:lambdafinal}
\end{equation}
and
\begin{equation}
\chi
=
\frac{a}
{\sqrt{(r^2+a^2-B)(r^2+a^2-f)}}.
\label{eq:chifinal}
\end{equation}

Introducing
\begin{equation}
\Gamma(r)
=
\sqrt{\frac{r^{2}+a^{2}-f(r)}
     {r^{2}+a^{2}-B(r)}},
\label{B32a}
\end{equation}
substitution of Eqs.~\eqref{eq:lambdafinal} and~\eqref{eq:chifinal} into Eqs.~\eqref{eq:gur} and~\eqref{eq:grphi} gives
\begin{equation}
g^{(\rm EF)}_{ur}
=
-\Gamma(r) ,
\qquad
g^{(\rm EF)}_{r\varphi}
=
a\,\Gamma(r)\sin^2\theta .
\label{eq:KSderived}
\end{equation}
It is noted that the off-diagonal component $g^{(\rm EF)}_{r\varphi}$ is not simply proportional to $a\sin^2\theta$, as in the Kerr case, but is dressed by the factor $\Gamma(r)$.
This is a direct consequence of the fact that the two metric functions $f(r)$ and $B(r)$ are distinct.
In the Kerr limit,
\begin{equation}
f(r)=B(r),
\end{equation}
one has
\begin{equation}
\beta(r)=1,
\qquad
\Gamma(r)=1,
\end{equation}
so that Eqs.~\eqref{eq:lambdafinal} and~\eqref{eq:chifinal} reduce to the standard Kerr coordinate transformation, and Eq.~\eqref{eq:KSderived} to the standard Kerr-Schild form.

Accordingly, the metric takes the generalized Kerr-Schild form
\begin{align}
ds^{2}
=&
-\left(
1-\frac{f(r)}{\Sigma}
\right)du^{2}
-2
\Gamma
\,du\,dr
+\Sigma\,d\theta^{2}
\nonumber\\
&
-\frac{2af(r)\sin^{2}\theta}{\Sigma}
\,du\,d\varphi
+2a\sin^{2}\theta
\Gamma
\,dr\,d\varphi
\nonumber\\
&
+\sin^{2}\theta
\left(
r^{2}+a^{2}
+\frac{a^{2}f(r)\sin^{2}\theta}{\Sigma}
\right)
d\varphi^{2}.
\label{B32}
\end{align}

The corresponding covariant metric tensor in the coordinate basis $(u,r,\theta,\varphi)$ is

\begin{equation}
g_{\mu\nu}
=
\begin{pmatrix}
-\left(1-\dfrac{f}{\Sigma}\right)
&
-\Gamma
&
0
&
-\dfrac{af\sin^{2}\theta}{\Sigma}
\\[2ex]
-\Gamma
&
0
&
0
&
a\sin^{2}\theta
\Gamma
\\[2ex]
0
&
0
&
\Sigma
&
0
\\[2ex]
-\dfrac{af\sin^{2}\theta}{\Sigma}
&
a\sin^{2}\theta
\Gamma
&
0
&
\sin^{2}\theta
\left(
r^{2}+a^{2}
+\dfrac{a^{2}f\sin^{2}\theta}{\Sigma}
\right)
\end{pmatrix}.
\label{eq:EFrot2}
\end{equation}

The inverse reads:
\newcommand{\Sg}{\Sigma}
\begin{equation}
g^{\mu\nu}=
\begin{pmatrix}
\dfrac{a^{2}\sin^{2}\theta}{\Sg} &
-\dfrac{r^{2}+a^{2}}{\Gamma\,\Sg} &
0 &
\dfrac{a}{\Sg}\\[12pt]
-\dfrac{r^{2}+a^{2}}{\Gamma\,\Sg} &
\dfrac{r^{2}+a^{2}-B}{\Sg} &
0 &
-\dfrac{a}{\Gamma\,\Sg}\\[12pt]
0 & 0 & \dfrac{1}{\Sg} & 0\\[12pt]
\dfrac{a}{\Sg} &
-\dfrac{a}{\Gamma\,\Sg} &
0 &
\dfrac{1}{\Sg\sin^{2}\theta}
\end{pmatrix},
\end{equation}

For completeness, the Newman-Penrose null tetrad reads:
\bqn
l^{\mu}&=&\delta^{\mu}_{r},\nb\\[4pt]
n^{\mu}&=&\frac{1}{\Gamma}\,\delta^{\mu}_{u}
-\frac{1}{2\Gamma^{2}}\left(1-\frac{f}{\Sigma}\right)\delta^{\mu}_{r},\nb\\[4pt]
m^{\mu}&=&\frac{1}{\sqrt{2}\,(r+ia\cos\theta)}
\left[\,ia\sin\theta\left(\delta^{\mu}_{u}-\frac{1}{\Gamma}\,\delta^{\mu}_{r}\right)
+\delta^{\mu}_{\theta}+\frac{i}{\sin\theta}\,\delta^{\mu}_{\phi}\right],\nb\\[4pt]
\bar{m}^{\mu}&=&\frac{1}{\sqrt{2}\,(r-ia\cos\theta)}
\left[-ia\sin\theta\left(\delta^{\mu}_{u}-\frac{1}{\Gamma}\,\delta^{\mu}_{r}\right)
+\delta^{\mu}_{\theta}-\frac{i}{\sin\theta}\,\delta^{\mu}_{\phi}\right].
\eqn

\subsection{Feasibility of modified Newman-Janis construction}

The discussion presented above shows that the Newman-Janis algorithm does not by itself determine the complexified functions $\widetilde A=\widetilde A(r,\theta)$ and $\widetilde G=\widetilde G(r,\theta)$.
Instead, these functions must satisfy the algebraic requirement that the contravariant metric reconstructed from the Newman-Penrose null tetrad, Eq.~\eqref{eq:ginvrot}, is the inverse of the covariant metric given by Eq.~\eqref{eq:EFrot2}, which manifestly satisfies the Einstein field equation. 
We now prove that this inverse-matrix condition cannot be satisfied, or in other words, no functions $\widetilde A=\widetilde A(r,\theta)$ and $\widetilde G=\widetilde G(r,\theta)$ will furnish Eq.~\eqref{Mansatz} to be a valid metric solution of the form proposed in this study.

Let us be more specific.
The inverse-matrix relation reads
\begin{equation}
g^{\mu\lambda}g_{\lambda\nu}
=
\delta^\mu_{\ \nu}.
\label{eq:inversecondition}
\end{equation}
At first sight Eq.~\eqref{eq:inversecondition} provides sixteen algebraic equations. 
However, because both the covariant and contravariant metrics possess the same sparse Kerr-like structure, most of these equations are either identically satisfied or linearly dependent.

For example, the $(u,r)$ component gives
\begin{equation}
g^{uu}g_{ur}
+
g^{u\phi}g_{\phi r}
=
0.
\end{equation}
Using Eqs.~\eqref{eq:ginvrot} and~\eqref{eq:EFrot2},
\begin{equation}
-\frac{a^2\Gamma\sin^2\theta}{\Sigma}
+
\frac{a}{\Sigma}
\left(a\sin^2\theta\Gamma\right)
=
0,
\end{equation}
which is an identity independent of $\widetilde A$ and $\widetilde G$.

Likewise, the $(r,\theta)$ component 
\begin{equation}
g^{\theta\lambda}g_{\lambda r}
=
0
\end{equation}
is also identically satisfied, while the diagonal component immediately gives
\begin{equation}
g^{\theta\theta}g_{\theta\theta}
=
\frac1{\Sigma}\Sigma
=
1.
\end{equation}
Therefore, these equations contain no information regarding the unknown functions.

After eliminating all such identities, only a few independent algebraic equations remain. 
For simpler scenarios, the number of independent conditions is precisely two, which suffices to uniquely determine the two unknown functions, $\widetilde A$ and $\widetilde G$.
For the present case, however, we show explicitly that the system is overdetermined, so that no valid solution for $\widetilde A$ and $\widetilde G$ exists.
The first independent equation is obtained from the $(r,r)$ component,
\begin{equation}
g^{r\mu}g_{\mu r}
=
1.
\end{equation}
Substituting Eqs.~\eqref{eq:ginvrot} and~\eqref{eq:EFrot2},
\begin{equation}
\left(-\frac{1}{\widetilde{A}} - \frac{a^2 \sin^2\theta}{\Sigma}\right)(-\Gamma) + \left(\widetilde{G} + \frac{a^2 \sin^2\theta}{\Sigma}\right)(0) + 0 + \left(-\frac{a}{\Sigma}\right)\left(a\Gamma\sin^2\theta\right) = 1 ,
\end{equation}
in which the terms proportional to $a^2\Gamma\sin^2\theta/\Sigma$ cancel identically, so that it simplifies to
\begin{equation}
\widetilde A=\Gamma . %=\sqrt{\frac{r^{2}+a^{2}-f(r)}{r^{2}+a^{2}-B(r)}}.
\label{eq:Atilde_solution}
\end{equation}

The second independent equation follows from the $(r,u)$ component,
\begin{equation}
g^{r\mu}g_{\mu u}
=
0.
\end{equation}
Using Eqs.~\eqref{eq:ginvrot},~\eqref{eq:EFrot2} and~\eqref{eq:Atilde_solution},
\bqn
\left(-\frac{1}{\Gamma} - \frac{a^2 \sin^2\theta}{\Sigma}\right)\left(-1 + \frac{f}{\Sigma}\right) - \Gamma\left(\widetilde{G} + \frac{a^2 \sin^2\theta}{\Sigma}\right) + \frac{a^2 f \sin^2\theta}{\Sigma^2} = 0 ,
\label{eq:Gderive1}
\eqn
which gives
\begin{equation}
\widetilde G = \frac{\Sigma-f}{\Gamma^2\Sigma} + \frac{a^2\sin^2\theta}{\Sigma}\frac{1-\Gamma}{\Gamma}.
\label{eq:Gtilde_solution}
\end{equation}

At this point, as $\widetilde A$ and $\widetilde G$ are entirely determined, one must expect that Eqs.~\eqref{eq:Atilde_solution} and~\eqref{eq:Gtilde_solution} are consistent with all remaining nontrivial components of Eq.~\eqref{eq:inversecondition}.
Unfortunately, this is not the case.
Let us consider the $(u,\phi)$ component, which gives
\bqn
g^{u\lambda}g_{\lambda\phi} = -\frac{a\Gamma\sin^2\theta}{\widetilde{A}} - \frac{a^3\Gamma\sin^4\theta}{\Sigma} + \frac{a(r^2+a^2)\sin^2\theta}{\Sigma} = 0 .
\label{eq:uphicond}
\eqn
Substituting Eq.~\eqref{eq:Atilde_solution} and making use of $\Sigma=r^2+a^2-a^2\sin^2\theta$, the left-hand side of Eq.~\eqref{eq:uphicond} is evaluated in closed form,
\bqn
g^{u\lambda}g_{\lambda\phi}
=
\frac{a^3\sin^4\theta\left(1-\Gamma\right)}{\Sigma} ,
\label{eq:uphiresidual}
\eqn
which vanishes only in the Kerr limit $\Gamma=1$, namely $f=B$, or in the static limit $a=0$.
For a generic dark matter profile with $f(r)\neq B(r)$, the residual in Eq.~\eqref{eq:uphiresidual} is nonvanishing and Eq.~\eqref{eq:uphicond} is violated.
We emphasize that this obstruction involves $\widetilde A$ alone and is entirely independent of $\widetilde G$, so that it cannot be circumvented by any choice of the latter.

Therefore we conclude that the proposed metric differs conceptually from the original Newman-Janis prescription and its alternatives. 
In the original Newman-Janis algorithm, the complexification is introduced at the level of the radial coordinate, so that the metric functions are first promoted to functions of $(r,\bar r)$ before the complex coordinate transformation is carried out. 
Consequently, the resulting rotating geometry generally depends on the particular complexification prescription adopted for the radial functions, and different prescriptions may lead to inequivalent rotating metrics.
By contrast, the present construction does not assume any {\it a priori} prescription for the complexified functions. 
Effectively, the Newman-Penrose construction can be viewed in terms of two undetermined functions, ${\widetilde A}(r,\theta)$ and ${\widetilde G}(r,\theta)$, in the inverse metric Eq.~\eqref{eq:ginvrot}, in the sense that these functions would be obtained from some specific complexification scheme applied term by term to the Taylor expansion of the original metric functions $A$ and $G$.
Nonetheless, the specific form of these two functions must be determined through the metric Eq.~\eqref{eq:EFrot2}, by requiring Eq.~\eqref{eq:inversecondition}.
The above derivation shows that this is not possible.
For the very same reason, the present metric does not belong to the class of metrics obtained using the alternative Newman-Janis algorithm proposed by Azreg-A{\"i}nou~\cite{agr-bh-Kerr-55}.

\bibliographystyle{h-physrev}
\bibliography{references_qian}

\end{document}